\documentclass[a4paper,fleqn,usenatbib]{mnras}

\usepackage{newtxtext,newtxmath}

\usepackage[T1]{fontenc}
\usepackage{ae,aecompl}

\usepackage{graphicx}	
\usepackage{amsmath}	
\usepackage{amssymb}	
\usepackage[export]{adjustbox}
\usepackage{natbib}
\usepackage{gensymb}
\usepackage{tikz}
\usepackage{amsmath}
\usepackage{multirow}
\usetikzlibrary{decorations.markings}

\DeclareMathAlphabet{\mathsc}{OT1}{cmr}{m}{sc}
\def\testbx{bx}%
\DeclareRobustCommand{\ion}[2]{%
	\relax\ifmmode
	\ifx\testbx\f@series
	{\mathbf{#1\,\mathsc{#2}}}\else
	{\mathrm{#1\,\mathsc{#2}}}\fi
	\else\textup{#1\,{\mdseries\textsc{#2}}}%
	\fi}
\newcommand\abs[1]{\left|#1\right|}
\newcommand{\HI}{\ion{H}{i}}
\newcommand{\HII}{\ion{H}{ii}}

\newcommand{\FeII}{\ion{Fe}{ii}}
\newcommand{\FeIII}{\ion{Fe}{iii}}
\newcommand{\MgI}{\ion{Mg}{i}}
\newcommand{\MgII}{\ion{Mg}{ii}}

\newcommand{\CaII}{\ion{Ca}{ii}}

\newcommand{\NII}{\ion{N}{ii}}

\newcommand{\OII}{\ion{O}{ii}}
\newcommand{\OIII}{\ion{O}{iii}}

\newcommand{\kms}{km s$^{-1}$}

\newcommand{\msun}{M$_\odot$}

\title[Observational signatures for cold-flow accretion]{Observational signatures of a warped disk associated with cold-flow accretion \thanks{Based on data obtained under the ESO programme 096.A-0303 at the European Southern Observatories with MUSE at the 8.2 m telescopes operated at the Paranal Observatory, Chile.}}

\author[Hadi Rahmani et al.]{
Hadi Rahmani,$^{1,2}$\thanks{E-mail: hadi.rahmani@gmail.com}
C\'{e}line P\'{e}roux,$^{1}$
Ramona Augustin,$^{1,3}$
Bernd Husemann,$^{4}$
\newauthor Glenn G. Kacprzak,$^{5}$
Varsha Kulkarni,$^{6}$
Bruno Milliard,$^{1}$
Palle M\o{}ller,$^{3}$
Max Pettini,$^{7}$
\newauthor Lorrie Straka,$^{8}$
Jo\"{e}l Vernet,$^{3}$
and Donald G. York$^{9,10}$
\\
$^{1}$Aix Marseille Universit\'{e}, CNRS, LAM (Laboratoire d'Astrophysique de Marseille) UMR 7326, 13388, Marseille, France\\
$^2$GEPI, Observatoire de Paris, PSL Research University, CNRS, Place Jules Janssen, 92190 Meudon, France\\
$^{3}$European Southern Observatory, Karl-Schwarzschildstrasse 2, D-85748 Garching bei M{\"u}nchen, Germany\\
$^{4}$Max-Planck-Institut f{\"u}r Astronomie, K{\"o}nigstuhl 17, D-69117 Heidelberg, Germany\\
$^{5}$Swinburne University of Technology, Victoria 3122, Australia\\
$^{6}$Department of Physics and Astronomy, University of South Carolina, Columbia, SC 29208, USA\\
$^{7}$Institute of Astronomy, University of Cambridge, Madingley Road, Cambridge CB3 0HA, UK\\
$^{8}$Sterrewacht Leiden, Leiden University, PO Box 9513, NL-2300 RA Leiden, the Netherlands\\
$^{9}$Department of Astronomy and Astrophysics, The University of Chicago, Chicago, IL 60637, USA\\
$^{10}$Enrico Fermi Institute, The University of Chicago, Chicago, IL 60637, USA
}

\date{Accepted XXX. Received YYY; in original form ZZZ}

\pubyear{2015}

\begin{document}
\label{firstpage}
\pagerange{\pageref{firstpage}--\pageref{lastpage}}
\maketitle
\begin{abstract}
We present MUSE observations of the field of the quasar Q0152$-$020 whose spectrum shows a Lyman limit system (LLS) at redshift $z_{\rm abs} = 0.38$, with a metallicity Z\,$\gtrsim 0.06$\,Z$_\odot$. The low ionization metal lines associated with the LLS present two narrow distinct absorption components with a velocity separation of $26$\,\kms. We detect six galaxies within 600\,\kms\ from the absorption redshift; their projected distances from the quasar sightline range from 60 to 200\,kpc. The optical spectra of five of these galaxies exhibit prominent nebular emission lines, from which we deduce extinction-corrected star formation rates in the range SFR = 0.06--1.3\,M$_\odot$~yr$^{-1}$, and metallicities between 0.2\,Z$_\odot$ and Z$_\odot$. The sixth galaxy is only detected in the stellar continuum. By combining our data with archival Keck/HIRES spectroscopy of the quasar and HST/WFPC2 imaging of the field, we can relate absorption line and galaxy kinematics; we conclude that the LLS is most likely associated with the galaxy closest to the quasar sight-line (galaxy ``a''). Our morphokinematic analysis of galaxy ``a'' combined with the absorption line kinematics supports the interpretation that one of the absorption components originates from an extension of the stellar disk of galaxy ``a'', while the other component may arise in accreting gas in a warped disk with specific angular momentum $\sim 3$ times larger than the specific angular momentum of the galaxy halo. Such warped disks are common features in hydrodynamical simulations of cold-flow accretion onto galaxies; the data presented here provide observational evidence in favour of this scenario. 
\end{abstract}

%
%
%
\section{Introduction}
Galactic growth is regulated by the mass and momentum exchange of galaxies with their surroundings. Galaxies acquire their mass and momentum dominantly via intergalactic medium (IGM) gas accretion and lose them via strong outflows driven by active galactic nuclei or supernovae. Such strong outflows, also referred to as feedback processes, are observed to be associated with galaxies over a broad range of mass and morphology and a wide range of redshift \citep{Lehnert96,Heckman00,Pettini01lbg,Martin05,Weiner09,Nestor11,Martin12,Erb15}. Moreover, feedback processes are the key input for the cosmological simulations to be able to reproduce the observed mass and luminosity distribution of galaxies \citep{Springel05,Sijacki07,Booth09,Oppenheimer10,Haas13,Vogelsberger14_1,Schaye15}. On the other hand, IGM gas accretion is inferred via a few indirect arguments: (i) The available cold gas reservoirs in star-forming galaxies at any redshift are hardly enough to sustain their star formation rates (SFR) over a time scale of a few Gyr \citep{Genzel10,Daddi10,Kennicutt12,Tacconi13,Leroy13,Scoville16}; (ii) The observed metallicity distribution of old stars in the Milky Way is inconsistent with that produced by 'closed-box' galaxy evolutionary models, and requires the inclusion of inflow of low metallicity gas (the so-called 'G-dwarf problem'; see \citealt{Larson72,Lynden-Bell75,Pagel75,Fenner03,Chiappini09}). (iii) The cosmic density of neutral hydrogen, the fuel for forming stars, \citep{Peroux03,Prochaska05,Noterdaeme09dla,Noterdaeme12dla,Zafar13} evolves very little compared to the cosmic density of the star formation rate \citep{Madau14}.

From a theoretical point of view, IGM gas filaments inflowing into a dark matter halo, at the virial radius, get shock heated to the virial temperature which is $\sim10^6$ K for a $L^*$ galaxy \citep{Silk77,Rees77}. Subsequently, this gas cools and sinks to the center of the halo where it can form stars \citep{Maller04}. However, more recent works in cosmological hydrodynamical simulations emphasize the importance of the filamentary nature of gas accretion onto galaxies, particularly at high redshifts where cosmic filaments are much denser than those in the local Universe \citep{Keres05,Dekel06,Brooks09,Faucher-Giguere11,van-de-Voort15}. Such a filamentary stream of gas can be dense enough to have a cooling time scale shorter than the compression time scale \citep{Binney77,Birnboim03}. Therefore, shocks are not formed to heat the gas to the virial temperature. This leads to the penetration of the filamentary cold gas ($T\sim10^4$ K) streams into the inner galactic region of the dark matter halo of galaxies. This mode of gas accretion is referred to as the ``cold mode'' accretion. 

The inflowing gas into a galaxy via the ``cold mode'' accretion is expected to directly reach the outskirts of the galaxy, delivering fuel for star formation and also angular momentum \citep{Shen06,Agertz09,Stewart13,Danovich15}. Accreted ``cold mode'' gas is then expected to corotate with the central disk, though with a higher angular momentum, forming a warped extended gaseous structure in the circumgalactic medium (CGM).

Warped extended disks inclined up to $\sim 20$\degree\ with respect to the stellar disk having radii in a range of a few kpc to larger than 100 kpc are frequently detected in \HI\ 21-cm observations in the local Universe \citep{Briggs90,Shang98,Garcia-Ruiz02,Heald11}. At higher redshifts, warped extended disks should be detectable with distinguished kinematics in absorption against background quasars \citep{Steidel02,Kacprzak10a,Bouche13,Bouche16,Peroux17,Ho17}. Such absorption lines are predicted to have velocity offset of $\sim100$ \kms\ with respect to the galaxy's systemic velocity in the same direction as galaxy rotation \citep{Stewart11b}. Therefore, the technique of quasar absorption line spectroscopy can be highly effective in searching for signatures of ``cold mode'' accretion (provided one avoids regions along the minor axis of a galaxy where galactic outflows may be taking place).

\MgII\ absorbers are amongst the best tracers of the cold gas associated with galaxies. Galactic winds and inflowing gas or a combination of both are postulated as the origin of the \MgII\ absorbers \citep{Bouche07_mgii,Chen10,Nestor11,Bordoloi11,Kacprzak12,Nielsen15}. We note that all \MgII\ absorbers are associated with Lyman limit systems (LLS) that are absorbers with N(\HI)$\gtrsim10^{17}$\,cm$^{-2}$. Several studies have shown that the \MgII\ equivalent width is dependent on the apparent position of the background source with respect to the host galaxy absorbers \citep[e.g.,][]{Steidel94ApjL,Bordoloi11,Bouche12a,Kacprzak12,Bordoloi14a,Nielsen15}. As an example, using a sample of 88 quasar-galaxy pairs, \citet{Kacprzak12} found that the covering fraction of \MgII\ absorbers is enhanced by 20--30\% along both the major and minor axes of the host galaxies. The authors suggested that the enhanced covering fraction close to the major axes is most likely due to the accreting gas coplanar with the galaxy disks while the enhanced covering fraction close to the minor axes is likely a signature of outflowing material.  

LLSs provide more robust tools for CGM studies as they are detected via their \HI\ absorption lines rather than metals. This is important for the study of IGM gas accretion as such a pristine gas even if missed in a search via metal absorption lines would still likely to be detected via its \HI\ absorption lines. Recently, there have been a few studies demonstrating a bimodality in the metallicity distribution of LLS at $z\lesssim1$ \citep{Lehner13,Quiret16,Wotta16}. It has been postulated that the low metallicity portion of this bimodal distribution may be due to IGM gas inflowing onto galaxies. Using cosmological hydrodynamic simulations, \citet{Hafen16} demonstrated that LLS at $z\lesssim1$ can be associated with cool gas inflows from the IGM as well as galactic winds. However, the authors did not find signatures of bimodality in the metallicity distribution of LLS. While this result is in contrast with the observations, it is worth noting that the discrepancy could be partly explained by the lack of low metallicity LLS in the simulations. However, confirming this scenario has to await the detection and detailed study of LLS host galaxy absorbers.  

Galactic inflows and outflows can be traced more directly from the absorption spectra of the galaxies themselves \citep[e.g.,][]{Steidel10,Rubin12,Martin12,Kacprzak14}. The metal absorption lines redshifted (blueshifted) with respect to the systemic velocity of the galaxies can trace the inflowing (outflowing) gas in this technique. \citet{Rubin12} detected redshifted metal absorption lines to be associated with $\sim6$\% of the galaxies in their sample. They further found that the majority of such galaxies are edge-on. \citet{Martin12} reported a similar detection rate of $\sim4$\% for redshifted absorption lines although they found no obvious correlation with the inclination of the galaxies in their sample. While this technique is simple and straightforward, it suffers from the lack of information about the apparent position of the absorbing gas and the low spectral resolution of such observations. 

Integral field unit (IFU) observations of quasar-galaxy pairs provide a powerful tool for probing the CGM. Such IFU data, combined with high resolution spectra of background quasars, make it possible to disentangle the origin of the gas detected in absorption \citep[e.g.,][]{Bouche07_simple,Peroux11a,Peroux13,Bouche13,Schroetter15,Peroux17}. \citet{Bouche13} studied a quasar sightline having a strong \MgII\ absorption at  $z=2.33$ with the host galaxy absorber at an impact parameter of 26 kpc. In a detailed analysis of the host galaxy's kinematics along with quasar absorption line kinematics, the authors found signatures of an extended cold gaseous disk consistent with that predicted in ``cold mode'' accretion. \citet{Bouche16} and \citet{Peroux17} reported further evidence of such extended gaseous disks based on studies of quasar-galaxy pairs at $z \lesssim 1$. Lyman$\alpha$ emission have been detected from the cosmic web in the vicinity of bright quasars \citep{Weidinger05,Cantalupo14,Martin-C15,Borisova16,Arrigoni-Battaia16ApJ}. Possible signatures of inflowing gas from cosmic web have been obtained using IFU observations of such objects \citep[e.g.,][]{Martin-C15,Martin16ApJ}.   

With this work, we add to this picture by reporting observational evidence for another warped galaxy disk which may be associated with cold-flow accretion. We present a study of galaxies in the field of the quasar Q0152$-$020, whose spectrum shows a LLS, using a combination of our own VLT/MUSE IFU observations together with archival Keck/HIRES and HST/FOS spectroscopy of the quasar and HST/WFPC2 imaging of the field. The paper is organized as follows. In Section (2) we describe all the available observations of this quasar field. In Section (3) we present the analysis of the absorption line spectra. In Section (4) we identify the absorbing galaxies from VLT/MUSE and study their star formation rate, metallicity and morphokinematics. We probe the origin of this absorption system in section (5) and summarize the results in Section (6).
%
%
\section{Observations of the field of Q0152$-$020}
Q0152$-$020 (or UM 675) is a bright quasar with a broad band magnitude of V = 17.4 mag \citep{Veron-Cetty10}. In this section we review the  spectroscopic and imaging observations of this quasar field that are relevant to the current study.
\subsection{Quasar spectroscopy}
\citet{Rao06} observed this quasar using the Faint Object Spectrograph (FOS) on Hubble Space Telescope (HST) at a spectral resolution of $FWHM\sim350$ \kms\ to measure the N(\HI) associated with the strong \MgII\ absorber at $z\sim0.78$. This HST/FOS spectrum covers a wavelength range from 1600 \AA\ to 3200 \AA\ which includes Lyman-$\alpha$ absorption from the $z =0.38$ system we consider in this paper. We obtained the 1D reduced HST/FOS quasar spectrum from MAST\footnote{https://mast.stsci.edu} dataset. 

Q0152$-$020 has also been observed using Keck/HIRES (PI: Tytler, Program ID: U031Hb). This high spectral resolution data (R $\equiv\Delta\lambda/\lambda$=48000) covers a wavelength range from 3245 \AA\ to 5995 \AA. We obtained the reduced normalized Keck/HIRES data of this quasar from the Keck Observatory Database of Ionized Absorbers towards quasars \citep[KODIAQ:][]{OMeara15}. 
\subsection[short title]{HST imaging}
We also obtained the Hubble Space Telescope (HST) WFPC2/$F702W$ imaging of this quasar field (PI: Steidel, Program ID: 6557) from the Hubble Legacy Archive \footnote{http://hla.stsci.edu} which provides enhanced HST products. This image has a $150"\times150"$ field of view with a $0.1"$ pixel scale where the quasar has $\sim30"$ offset with respect to the center of the field. Using a few point sources in the field, we measure a spatial resolution of $FWHM=0.2"$ for this image. 
\subsection[MUSE observation]{New MUSE observations}
We have obtained IFU observations of the field of Q0152$-$020 using VLT/MUSE under the ESO program 96.A-0303 (PI: P\'{e}roux). The observations were carried out in service mode in seeing limited mode. The total of 100 minutes exposure time was achieved using two observing blocks (OBs) each 50 minutes. The field was rotated by 90 degrees between the two OBs to optimize the flat fielding. For details of the observing strategy and data reduction steps, we refer the reader to \citet{Peroux17}. In a nutshell, the data are reduced with version v1.6 of the ESO MUSE pipeline \citep{Weilbacher15}. We check that the flat-fields are the closest possible to the science observations in terms of ambient temperature to minimize spatial shifts. The raw science data are then processed with $scibasic$ and then $scipost$ recipe is run with the sky subtraction method ``simple'' on. During this step, the wavelength calibration is corrected to a heliocentric reference. The individual exposures are registered using the point sources in the field within the $exp\_align$ recipe, ensuring accurate relative astrometry. Finally, the individual exposures are combined into a single data cube using the $exp\_combine$ recipe. Following \citet{Peroux17}, the removal of OH emission lines from the night sky is accomplished with an external routine. After selecting sky regions in the field, we create ``\textit{principal component analysis}'' components from the spectra which are further applied to the science data cube to remove sky line residuals \citep{Husemann16}.

The resulting data cube covers emission lines from [\OII]\,$\lambda \lambda 3726, 3729$ to H$\alpha$ at $z=0.38$. The spectral resolution is R=1770 at 4800 \AA\ and R=3590 at 9300 \AA\ resampled to a spectral sampling of 1.25 \AA/pixel. The seeing of the final combined data has a Gaussian full width at half maximum of $0.70 \pm 0.02$ arc second ($\equiv3.5\pm0.1$ pixel) at 7000 \AA\ as measured using the quasar and two stars in the MUSE field of view. 

We check the wavelength solution using the known wavelengths of the night-sky OH lines and found it to be accurate within 20 \kms. To test the accuracy of the flux calibration of the MUSE data, we measure the broad band AB magnitudes of 10 isolated objects over the MUSE field of view and compare them with those obtained from the HST/WFPC2 image. To do so we convolve the extracted 1D spectra of these objects with the response function of the $F702W$ filter and integrate to obtain the broad band magnitudes. We notice that the broad band fluxes of the MUSE objects are inconsistent by $\sim10$\% with respect to those from HST/WFPC2. To be conservative we inflate our statistical measured flux errors by 10\% to include this possible systematic error from flux calibration. 
\begin{figure}
	\centering	
	\includegraphics[width=0.9\hsize,bb=135 10 475 749,clip=,angle=0]{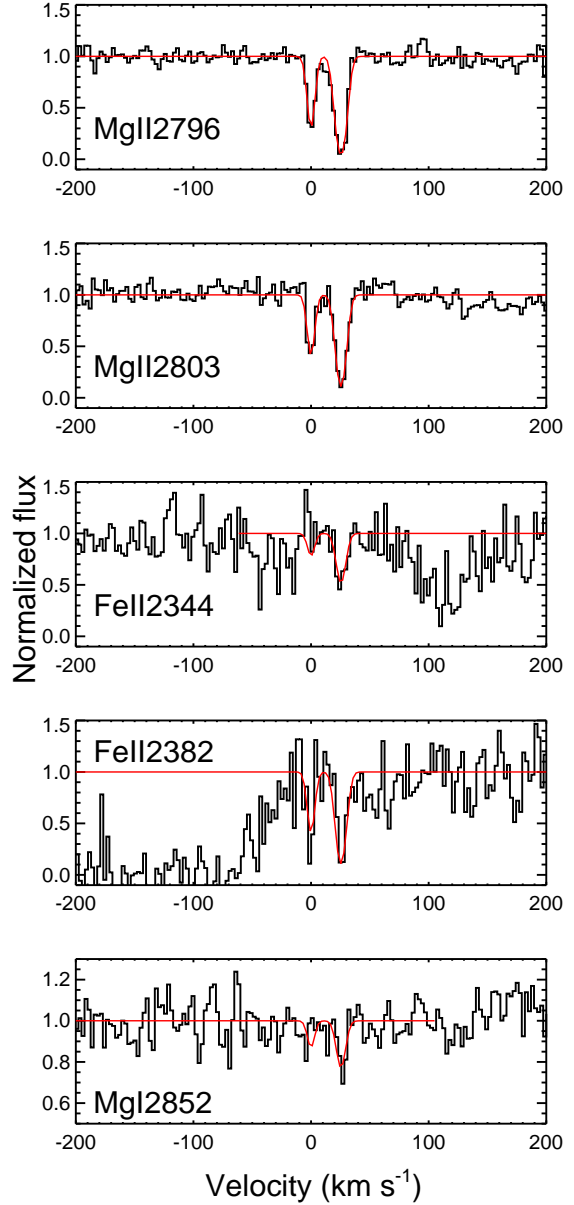}
	\caption{Normalized portions of the Keck/HIRES spectrum of Q0152$-$020. The observed spectrum and best Voigt profile fits are shown using histogram and continuous lines, respectively (note the different y-scale in the case of \MgI). The zero velocity is set at the redshift of the weaker metal absorption component at $z=0.38296$. The bluer \MgI\ component is consistent with a $3\sigma$ upper limit of $\log$ [N(\MgI)/cm$^{-2}$] $<10.9$. }
	\label{fig_mgii_zp3}
\end{figure}
\begin{figure*}
	\centering	
	\hspace*{-2.1cm}
	\hbox{
		\includegraphics[width=.65\textwidth,bb=18 180 594 612,clip=,angle=0]{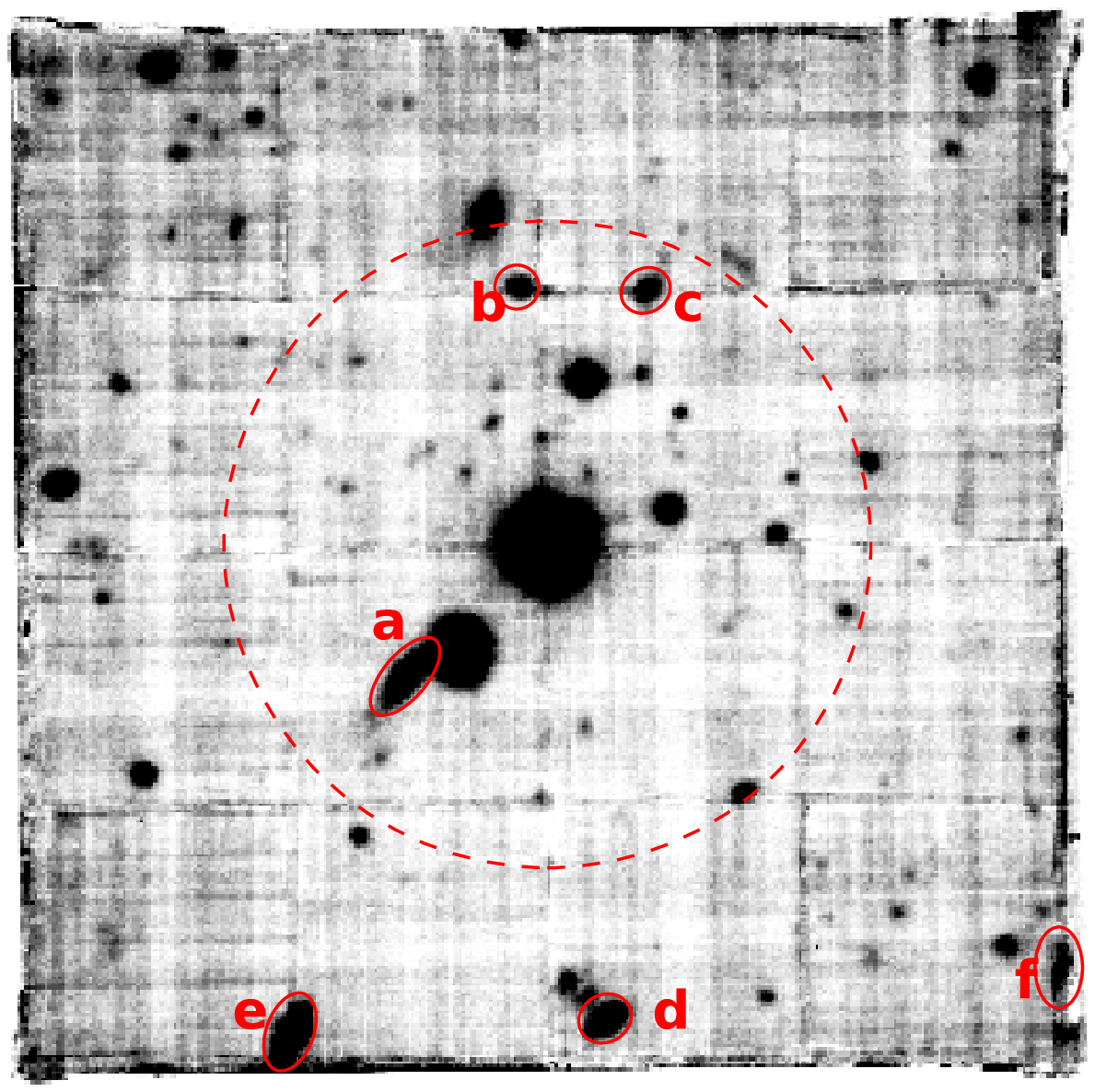} 
		\hspace*{-1.9cm}   	
		\includegraphics[width=.65\textwidth,bb=18 180 594 612,clip=,angle=0]{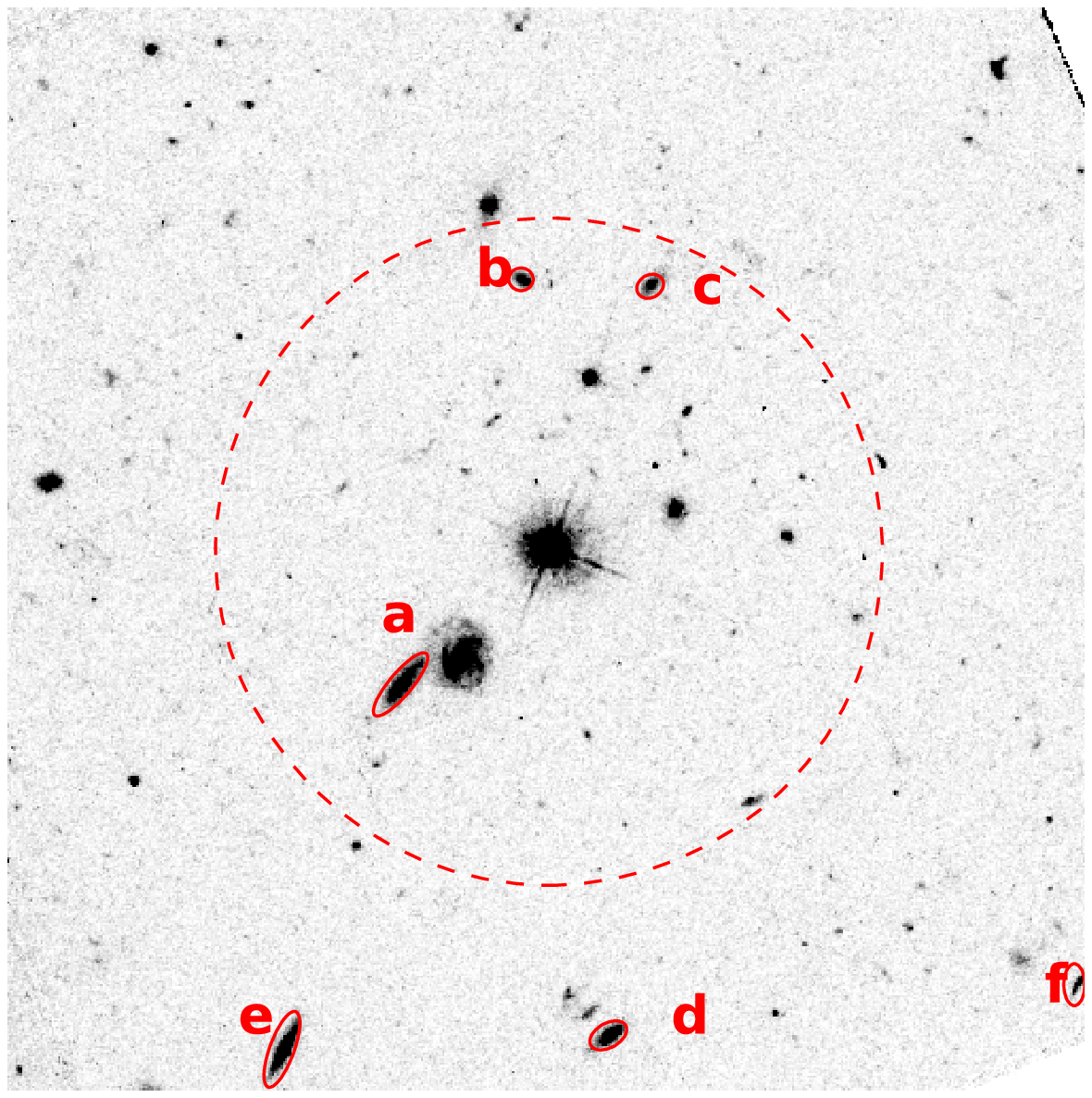}
	}
	\caption{{\it left}: The $1'\times1'$ white light image produced by collapsing the MUSE cube along the wavelength direction. The ``a'' to ``f'' letters mark the sky position of the galaxies at a mean redshift of $z=0.3816$, ordered by impact parameter, and the quasar resides in the centre of the image as indicated by \textbf{Q}. We have indicated these galaxies with ellipses. Galaxy ``e'' is detected only in continuum with no detectable emission line (see Fig. \ref{fig_Gs_spec}). {\it right}: The $1'\times1'$ HST/WFPC2 $F702W$ image of the same field. As indicated by ellipses and with the same names, all the 6 galaxies matching the absorption redshift are also detected in the HST image. North and East are towards up and left, respectively. A projected distance of 100 kpc at the absorption redshift is indicated by a dashed circle in each panel.}
	\begin{picture}(0,0)(0,0)
	\linethickness{.2mm}
	\put( 10,90){\textcolor{black}{\vector(0,1){35}}}	
	\put( 10,90){\textcolor{black}{\vector(-1,0){35}}}	
	\put(-0,130){\bf \large N}  \put( -40,90){\bf \large E}
	\put( -137,217){\textcolor{red}{\bf \large Q}}
	\put( 142,217){\textcolor{red}{\bf \large Q}}
	\end{picture}
	\label{fig_muse_white}
\end{figure*}
\begin{figure*}
	\centering	
	\centerline{\includegraphics[width=1.1\textwidth,bb=0 0 595 842,clip=,angle=90]{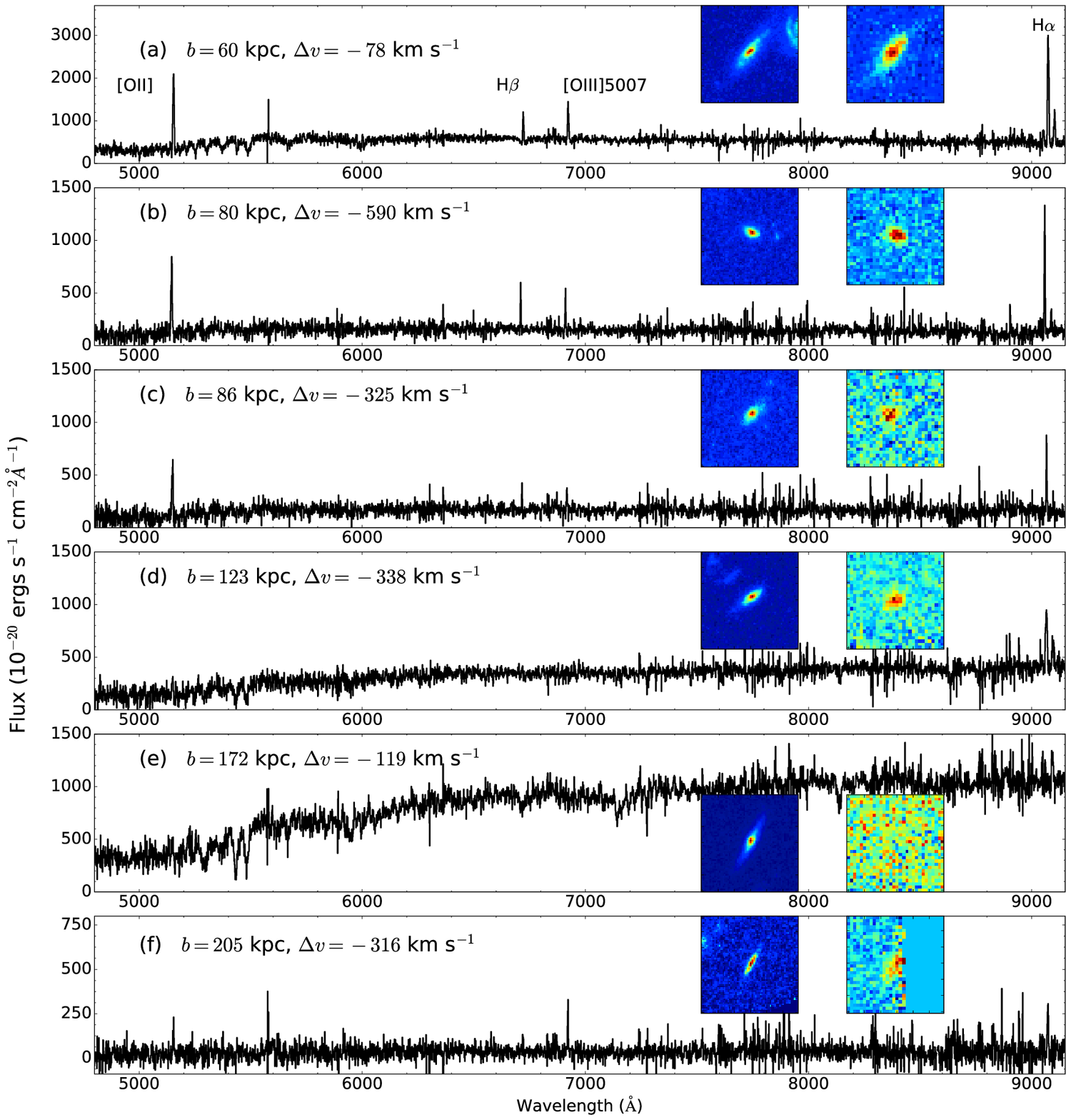}}
	\vspace*{-15mm}
	\caption{Extracted 1D MUSE spectra of the galaxies at $z=0.38$ in an order of increasing impact parameter. The $6"\times6"$ HST/WFPC2 $F702W$ and H$\alpha$ MUSE narrow band post stamp images of each galaxy are also shown on the top right of each panel. In the top left of each panel the impact parameter and velocity offset with respect to the absorber are given. The emission line at $\lambda\sim9090$ \AA\ seen in galaxy ``e'' is the [\OII]\,$\lambda\lambda$3726,3729 doublet from a galaxy at $z=1.44$.}
	\begin{picture}(0,0)(0,0)
	\put(-177,510){ \sffamily CaII H\&K} 
	\put( -150,509){\textcolor{black}{\vector(0,-1){20}}}	
	\put( -144,509){\textcolor{black}{\vector(0,-1){20}}}	
	\end{picture}
	\label{fig_Gs_spec}
\end{figure*}
\section[short title]{Quasar absorption line analysis}
\begin{table}
	\caption{Metal absorption line properties as extracted from the Voigt profile modeling of the Keck/HIRES spectrum of the quasar.}
	\begin{tabular}{lcccccc}
		\hline
		component& (1)&(2)\\
		\hline
		redshift&0.38296&0.38308\\
		b(\kms) &$1.9\pm0.4$&$3.8\pm0.3$\\
		$\log$[N(\MgI)/cm$^{-2}$]&$<10.9$&$11.3\pm0.1$\\
		$\log$[N(\MgII)/cm$^{-2}$]&$>12.7$&$>13.1$\\
		$\log$[N(\FeII)/cm$^{-2}$] &$12.5\pm0.2$ &$13.0\pm0.1$\\
		\hline
		\multicolumn{3}{l}{total column density of \MgII: log[N(Mg II)/cm$^{-2}$] $> 13.2$} \\
		\multicolumn{3}{l}{total column density of \FeII: log[N(Fe II)/cm$^{-2}$] $=13.1\pm0.1$} \\		
		\hline
	\end{tabular}
	\label{tab_all_in_absorption}
\end{table}
The Lyman-$\alpha$ absorption from this LLS at $z=0.38$ corresponds to an observed wavelength of $\sim1680$ \AA. Although this is covered by the HST/FOS spectrum, it resides close to the blue edge of the spectrum with a poor SNR. Given the quality of this spectrum we infer an upper limit of N(\HI)$<6\times10^{18}$ cm$^{-2}$ associated with this LLS (private communication, S. Rao).

Using the Keck/HIRES spectrum we measure the rest frame equivalent width of the \MgII$\lambda$2796 to be $0.17\pm0.01$ \AA. \MgII\ absorbers at such EWs (EW$\lesssim0.3$\AA) are considered as weak \MgII\ absorbers and known to be associated with optically thin gas at metallicities not less than 0.1 Z$_\odot$ \citep{Churchill99}. 

Fig. \ref{fig_mgii_zp3} presents \MgII, \FeII\ and \MgI\ absorption profiles where the continuous red lines show the best Voigt profile models obtained using the \textsc{vpfit}\footnote{http://www.ast.cam.ac.uk/$\sim$rfc/vpfit.html} v10.0 code. We have assumed a given component to have the same redshift and broadening parameter for all transitions. The absorption profiles consist of two very narrow components with velocity broadening parameters $b=1.9\pm0.4$ \kms\ and $3.8\pm0.3$ \kms. The two components have a velocity separation of $\Delta v = 26.0\pm0.3$ \kms. \MgI\ is not detected in the bluer component where we measure a $3\sigma$ upper limit of $\log$[N(\MgI) cm$^{-2}$] $<10.9$.  

From our best Voigt profile fit to the absorption lines we infer that both components of the \MgII\ doublets are saturated. Such unresolved saturations can be inferred also from the equivalent width ratio of the \MgII\ doublet which is $1.11\pm0.02$ and $1.07\pm0.01$ for the first and second components, respectively. This can be further appreciated by noting that it is possible to obtain a reasonable Voigt profile fit to the \MgII\ absorption lines by having $\sim10$ times larger column density for \MgII. In the latter case the $\chi^2$ of the fit increases only by $2$\%, with respect to our best model, which is negligible. On the contrary, we find the column density of \FeII\ to be stable in either of the fits. Therefore, to be conservative, we quote the \MgII\ column densities obtained from the best Voigt profile fits as lower limits. Table \ref{tab_all_in_absorption} summarizes the redshifts, broadening parameters and column densities for the achieved Voigt profile fit. 

Based on the upper limit on $\log$[N(\HI)/cm$^{-2}$]$<$18.8 we find a [Fe/H] ($\equiv$(Fe/H)-(Fe/H)$_\odot$) $>-1.2$ which can be considered as a lower limit of the metallicity \citep[(Fe/H)$_\odot=-4.5$,][]{Asplund09}. Indeed, our measured N(\HI) and metallicity are also in agreement with the expected properties of weak \MgII\ systems \citep{Churchill99,Rigby02}. \FeII\ is frequently detected to be depleted into dust in DLAs and sub-DLAs \citep{Vladilo98,Ledoux02a,Khare04,Wolfe05,Jenkins09}. The dust correction will increase the iron abundance and hence the metallicity lower limit still holds even in the presence of \FeII\ depletion. In addition, it is known that, a certain fraction of hydrogen and iron is in \HII\ and \FeIII\ phase at such \HI\ column densities. However, the ionized fractions are expected to be very close for \HI\ and \FeII\ to keep the [\FeII/\HI] ratio still as a valid tracer of [Fe/H] \citep[see][for example]{Dessauges-Zavadsky03}. We caution the reader that the lowest column density studied by \citet{Dessauges-Zavadsky03} to probe the ionization effects in LLS is roughly two times higher than the measured N(\HI) upper limit for this absorber. 

To study the absorption line kinematics we measure the quantity $\Delta v_{90}$ which is defined as the velocity range over which the integrated optical depth ($\tau_{tot}~ = ~ \int_{-\infty}^{+\infty} \tau(v) ~ dv$) reaches from 5\% to 95\% of its total value \citep{Prochaska97,Ledoux06a}. We measure the $\Delta v_{90}$ using the technique described in \citet{Quiret16} which uses the best fit Voigt profile model instead of the observed spectrum. Integrating over the noise and blend free Voigt profile obtained from a simultaneous fit of several absorption lines in this approach leads to a more robust estimate of $\Delta v_{90}$. We obtain a $\Delta v_{90}=30$ \kms\ using \FeII\,$\lambda$2344 line. 

\section[galaxy group in MUSE]{galaxies close to the absorption redshift}\label{group_muse}
%
Knowing the redshift of the \MgII\ absorber at $z=0.38$ we looked for galaxies at $\Delta v\lesssim600$ \kms\ and $b\lesssim200$ kpc with respect to the absorber in the MUSE data cube. We searched for such galaxies by inspecting the narrow-band images created using the MUSE data cube centered at the expected wavelengths of different nebular emission lines. As the result we found 5 galaxies close to the absorption redshift each with several emission lines. The impact parameters of these galaxies are in the range of 60 to 205 kpc. Moreover, all these galaxies are blueshifted with respect to the absorbing system at velocities from $-78$ to $-590$\,\kms. 

\citet{Rahmani16} have shown that the host galaxies of absorbing systems can be quenched galaxies with no detectable emission line \citep[see also][]{Chen10,Bielby17}. These galaxies are not detected in a search using narrow-band images centered around the expected wavelengths of the emission lines. To search for such objects we created a white light image from the MUSE data cube and execute \textsc{sextractor} \citep{Bertin96} to find all objects with continuum emission. In the spectra of galaxies brighter than $m_r\sim23$ mag and with no emission lines that are only detected in continuum we then looked for absorption diagnostics (\CaII\ H\&K and 4000\,\AA\ break, in particular) at $z\sim0.38$. As a result of this search, we found a quenched galaxy as the 6th member of this group at an impact parameter of $b=172$ kpc. By modeling \CaII\ H\&K absorption lines we find $z=0.38135\pm0.00014$ that shows this galaxy is blueshifted with respect to the absorber at $v=-119$ \kms. We label these galaxies as objects ``a'' thru ``f'' based on increasing impact parameters. 

In the \textit{left} panel of Fig. \ref{fig_muse_white} we indicate these galaxies with ellipses overlaid on the white light image generated from the MUSE data cube. The quasar is marked with a ``Q'' at the center of the field of view. In the \textit{right} panel of Fig. \ref{fig_muse_white} we present the HST/WFPC2 image of the same field of view. As can be inferred from HST/WFPC2 image all these six galaxies are detected in continuum.

To extract the spectrum of each galaxy from the MUSE cube we made use of the MUSE Python Data Analysis Framework  (\textsc{mpdaf v2.0}). This tool is a Python based package, developed by the MUSE Consortium \footnote{http://mpdaf.readthedocs.io/en/stable/credits.html}, for manipulating the MUSE data cubes. We extracted the 1D spectrum of each object through the following steps: (1) generating a subcube where the object of interest was placed in its center; (2) generating a 2D white light image by integrating the flux density at each pixel; (3) finding pixels associated with this object and those with the background using \textsc{sextractor}; (4) extracting the 1D spectrum of the object by integrating the flux of the pixels associated to this object in each wavelength plane. 

We find the flux of each emission line by modeling its profile with a Gaussian. We notice that modeling [\OII]\,$\lambda\lambda$3726,3729 emission usually requires a double Gaussian, although the doublet is only marginally resolved by the MUSE spectral resolution. For any of the non-detected emission lines we find the flux upper limit based on the flux $rms$ at the expected wavelength with a spectral $FWHM = 2.5$ \AA. 

Fig. \ref{fig_Gs_spec} presents the spectra of these 6 galaxies in the observed frame. We have marked the wavelengths of the strong nebular emission lines and \CaII\ H\&K absorption lines in the first panel. The $6"\times6"$ size narrow band H$\alpha$ map and HST/WFPC2 image of the same galaxy are shown in each panel. The spectrum of galaxy ``a'' presents strong Balmer absorption lines that trace a population of young and intermediate age stars. Galaxies ``b'' and ``c'' have fainter continua compared to galaxy ``a''. Galaxy ``d'' presents the broadest emission lines amongst these galaxies with a $FWHM=257$ \kms\ but marginally detected in [\OIII]\,$\lambda$5007. Strong absorption features of \CaII\ H\&K lines along with Balmer lines and Mgb triplet lines are detected in the spectrum of galaxy ``e''; these spectral features are typical of passive galaxies. It is worth noting that our flux upper limit for galaxy ``e'' translates to an upper limit on the {$\rm H\alpha$} luminosity of $<0.4\times10^{40}$ ergs s$^{-1}$. This value is of the order of (or smaller than) the {$\rm H\alpha$} luminosity of the Small Magellanic Cloud and an order of magnitude smaller than the {$\rm H\alpha$} luminosity of the Large Magellanic Cloud \citep{Kennicutt08}. We also detect an emission line at $\lambda\sim9090$ \AA\ in the spectrum of galaxy ``e''. However, the emission profile appears like a doublet where the wavelength separation matches a [\OII]\,$\lambda\lambda$3726,3729 emission from a galaxy at $z=1.44$. Galaxy ``f'' is marginally covered in the MUSE data and we detect multiple emission lines from this galaxy but a very faint continuum. 

We report the mean redshift obtained from all the detected emission lines as the redshift of each galaxy in Table \ref{tab_flux}. For the case of galaxy ``e'', with no detected emission line, we use the \CaII\ absorption lines to estimate the redshift. The total fluxes of different emission lines are also collected in Table \ref{tab_flux}. 

We use the luminosity function of H$\alpha$ emitters at $z=0.40$ \citep{Ly07} to calculate the comoving number density of H$\alpha$ emitters $n_{\rm gal}$. If we integrate from our detection limit of $0.4\times10^{40}$\,ergs\,s$^{-1}$ we find $n_{\rm gal}\cong0.025$ Mpc$^{-3}$. We then calculate the volume corresponding to the field of view of MUSE and the maximum velocity separation between two of the galaxies in the group (512 km s$^{-1}$ between galaxies ``a'' and ``b") to be 1.6 Mpc$^3$. Therefore, the expected number of H$\alpha$ emitters is 0.04. This is 125 times smaller than the number of detected H$\alpha$ emitters which is a sign of an overdensity of galaxies. We caution the reader that for accurately quantifying the overdensity one needs to integrate the galaxy correlation function over the volume defined by this group of galaxies. However, as this is beyond the scope of this work and also will not impact our conclusions we do not carry such an analysis. We use the \textit{gapper} estimator \citep[e.g.,][]{Yang05} to obtain the velocity dispersion of these galaxies, $\sigma_v=192$ \kms. 

In the rest of this section we present extracted properties of these galaxies based on MUSE and HST/WFPC2 data. We note that galaxy ``b'' at $\Delta v=-590$ \kms\ may not be related to the absorption system due to a large velocity separation. However, as this galaxy belongs to the same overdensity of galaxies we present its properties along with those of other galaxies for completeness.
\subsection[extinction]{Extinction correction}
Accurate H$\alpha$ and H$\beta$ flux measurements from each of these galaxies allow us to have precise estimates of their intrinsic dust attenuations. For a case B recombination at a temperature of $T=10^4$ K and an electron density of $n_e=10^2$--$10^4$ cm$^{-3}$, which is typical for \HII\ regions, the intrinsic value of the Balmer decrement is 2.88 \citep{Osterbrock89book}. To estimate the color excess we adopt a Small Magellanic Cloud-type (SMC-type) extinction law and utilize the following parametrization
\begin{equation}
E(B-V) = \frac{1.086}{k({\rm H\beta})-k({\rm H\alpha})} \log \left(\frac{\rm H\alpha}{\rm 2.88 H\beta}\right)
\end{equation}
where $k(\rm H\alpha)$ and $k(\rm H\beta)$ are respectively the extinction curve values at $\lambda$ = 6564 \AA\ and 4862 \AA\ taken from \citet{Pei92}. Based on the estimated values of $E(B-V)$ and extinction curves we correct the fluxes of all the emission lines for the intrinsic dust reddening. Table \ref{tab_flux_corrected} presents the $E(B-V)$ and dust corrected fluxes of all the emission lines. 
\subsection[SFR]{Star formation rates}
We convert the total H$\alpha$ extinction corrected fluxes of the galaxies to SFR assuming a \citet{Kennicutt98} conversion corrected to a \citet{Chabrier03} initial mass function. Our SFR measurements are presented in Table \ref{tab_params_emission}. The SFR of these galaxies are in the range of 0.06 -- 1.27 $M_\odot$ yr$^{-1}$ with a typical error of 0.01 M$_\odot$ yr$^{-1}$ (1$\sigma$). Furthermore, we use the upper limit on the {$\rm H\alpha$} flux of galaxy ``e'' to find a ${\rm SFR}<0.02$ M$_\odot$ yr$^{-1}$ (1$\sigma$). We recall that we have not fully covered galaxy ``f'' in our MUSE data and hence the measured SFR is a lower limit.

\begin{figure}
	\centering	
	\includegraphics[width=1.\hsize,bb=0 0 576 432,clip=,angle=0]{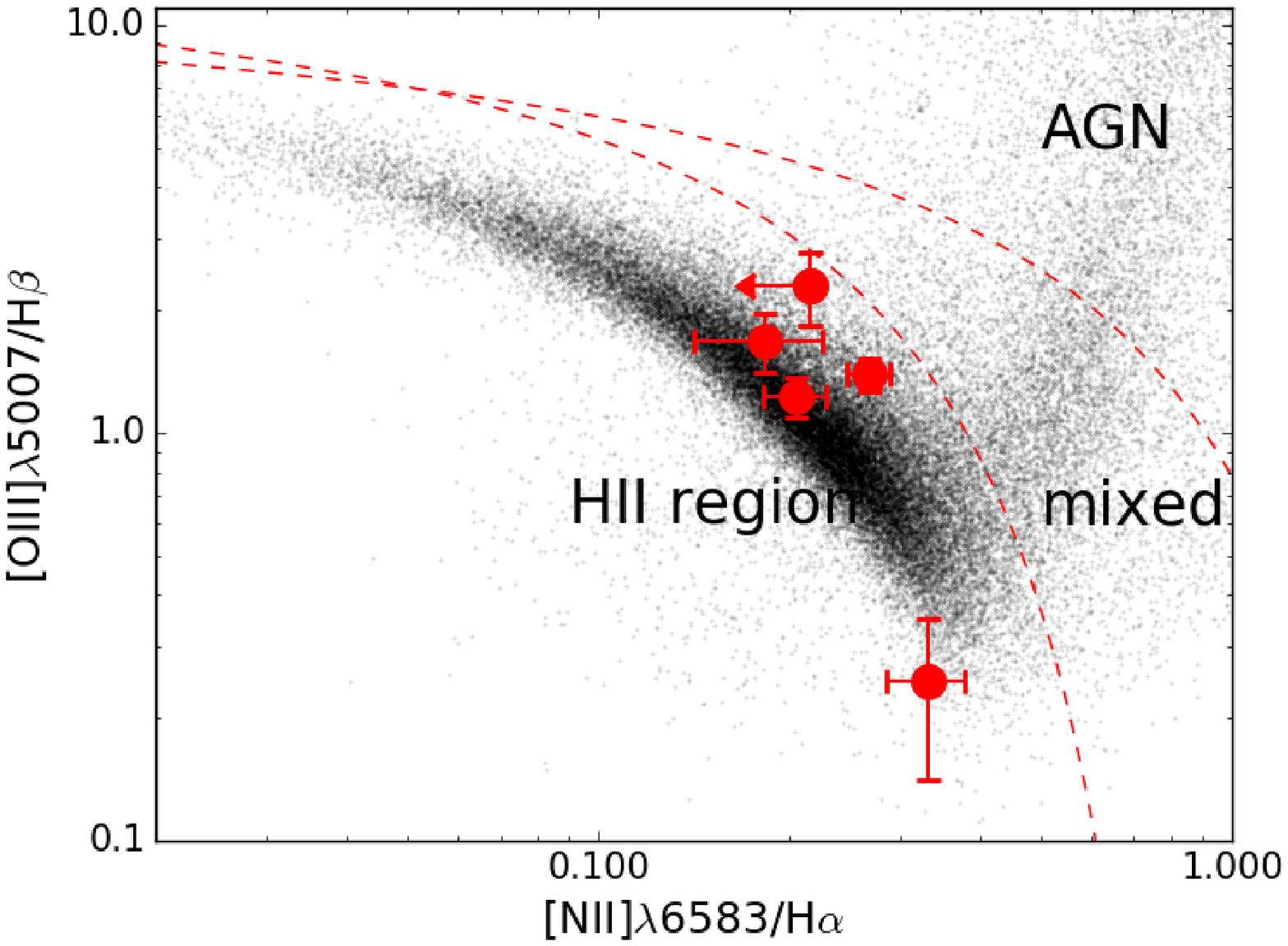}
	\caption{BPT diagram \citep{Baldwin81}. Black data points are the position of emitters in the SDSS DR8 \citep{Aihara11} where the emission line parameters were obtained from the MPA-JHU compilation \citep{Kauffmann03_sdss,Brinchmann04,Tremonti04}. The dashed lines mark the boarders between star forming galaxies and AGN \citep{Kewley01,Kauffmann03}. Red circles indicate the positions of the five emission line galaxies considered in this paper; all of them reside in the star forming region of the BPT diagram.}
	\label{fig_bpt}
\end{figure}

In converting the H$\alpha$ to SFR we have assumed that the \HII\ region is ionized by the Lyman limit photons from the young stellar objects and there is no contribution from active galactic nuclei (AGNs). We investigate the possible contribution of AGN to the fluxes of the emission lines by utilizing the BPT diagram \citep*[][]{Baldwin81}. Fig. \ref{fig_bpt} presents the BPT diagram where the abscissa and ordinate are respectively flux ratios of [\OIII]\,{$\rm \lambda$}5007/H$\beta$ and [\NII]\,{$\rm \lambda$}6853/H$\alpha$. Dashed lines, obtained from a combination of theoretical calculations and using galaxies contained in the Sloan Digital Sky Survey Data Release 1 \citep{Kewley01,Kauffmann03}, broadly define the separation between emission lines originating from star forming regions and AGN activity. A mixed nature of ionizing spectra is expected from objects occupying the region in between the two dashed lines. The 5 galaxies with detected emission lines are shown in the BPT diagram using filled circles. The black dots present the positions of galaxies and AGNs in SDSS DR8 \citep{Aihara11} where the line measurements were obtained from a MPA-JHU\footnote{http://wwwmpa.mpa-garching.mpg.de/SDSS/} run \citep{Kauffmann03_sdss,Brinchmann04,Tremonti04}. Clearly our 5 galaxies reside in the star forming region of the BPT diagram. Hence, systematic errors in the measured SFRs introduced by AGN contamination are thought to be negligible.
\begin{table*}
	\small
	\caption{Measured fluxes of nebular emission lines for galaxies in unit of $10^{-17}$ ergs s$^{-1}$ cm$^{-2}$.}
	\begin{tabular}{lccccccccccccc}
		\hline
		ID &  redshift$^1$ &  [\OII] & H$\beta$ & [\OIII]\,$\lambda$4958 & [\OIII]\,$\lambda$5007  &  [\NII]\,$\lambda$6583 & H$\alpha$ \\
		\hline
		a & $0.38260\pm0.00006$  &  $14.2\pm0.6$ & $4.2\pm0.3$ & $1.5\pm0.2$ & $6.2\pm0.3$   &$5.4\pm0.4$ & $20.3\pm0.4$ \\
		b & $0.38024\pm0.00003$  & $4.6\pm0.4$   & $1.5\pm0.1$ & $0.6\pm0.1$ & $1.8\pm0.2$  &$1.2\pm0.1$& $5.7\pm0.2$ \\
		c & $0.38146\pm0.00005$  & $4.3\pm0.6$   & $0.8\pm0.1$ & $0.4\pm0.1$ & $1.4\pm0.1$  &$0.6\pm0.1$& $3.3\pm0.2$   \\
		d & $0.38140\pm0.00024$  & $1.7\pm0.5$   & $1.0\pm0.1$ & $<0.2$ & $0.3\pm0.2$  &$1.7\pm0.2$& $5.1\pm0.2$ \\
		e  & $0.38135\pm0.00014$ & $<0.7$ & $<0.3$& $<0.4$& $<0.8$& $<1.1$ & $<0.8$\\
		f$^2$ & $0.38241\pm0.00004$  & $1.0\pm0.2$   & $0.5\pm0.1$ & $0.3\pm0.1$ & $1.2\pm0.1$ & $<0.4$& $1.8\pm0.2$ \\		
		\hline
	\end{tabular}
	\label{tab_flux}
	\begin{flushleft}
		$^1$ Redshifts are reported as the mean redshift of all detected emission lines with the exception of galaxy ``e'' where we measure the redshift from \CaII\ H\&K absorption lines.\\
		$^2$ This object is only partially covered by the MUSE field. Hence all measured fluxes should be considered as lower limits. The quoted errors are purely statistical. \\
	\end{flushleft}
\end{table*}
\begin{table*}
	\small
	\caption{Reddening corrected fluxes of nebular emission lines ( in unit of $10^{-17}$ ergs s$^{-1}$ cm$^{-2}$) based on the E(B$-$V) values reported in the last column.}
	\begin{tabular}{lccccccccccccc}
		\hline
		ID &    [\OII] & H$\beta$ & [\OIII]\,$\lambda$4958 & [\OIII]\,$\lambda$5007  &  [\NII]\,$\lambda$6583 & H$\alpha$  & E(B$-$V)$^1$\\
		\hline
		a &   $109.6\pm4.0$ & $18.9\pm1.5$ & $6.5\pm1.0$ & $26.3\pm1.2$   &$14.6\pm1.1$ & $54.6\pm1.3$ & $0.49\pm0.07$\\
		b & $16.7\pm1.5$   & $3.7\pm0.3$ & $1.5\pm0.3$ & $4.6\pm0.4$  &$2.2\pm0.2$& $10.7\pm0.3$  & $0.31\pm0.07$ \\
		c &  $19.0\pm2.8$   & $2.4\pm0.3$ & $1.2\pm0.3$ & $4.0\pm0.4$  &$1.3\pm0.3$& $6.8\pm0.4$ & $0.36\pm0.12$  \\
		d &  $16.6\pm4.6$   & $5.4\pm0.6$ & $<1.2$ & $1.3\pm0.9$  &$5.1\pm0.7$& $15.5\pm0.7$& $0.54\pm0.10$ \\
		f & $2.3\pm0.5$   & $0.9\pm0.2$ & $0.6\pm0.2$ & $2.1\pm0.2$ & $<0.6$& $2.6\pm0.2$ & $0.19\pm0.17$\\		
		\hline
	\end{tabular}
	\label{tab_flux_corrected}
	\begin{flushleft}
		$^1$The reddening is in unit of mag.
	\end{flushleft}
\end{table*}
\begin{table*}
	\small
	\caption{Estimated physical parameters of the galaxies. The absorption metallicity of Z$_{\rm abs}\gtrsim-1.2$ translates to 12+(O/H)$_{\rm abs}$ $\gtrsim$ 7.5. }
	\begin{tabular}{cccccccccccccc}
		\hline
		ID & RA & DEC&  $\Delta v^{1}$ & impact    & SFR& 12+$\log$(O/H) & 12+$\log$(O/H) & $M_{\rm HST}$\\
		&J2000&J2000&[\kms]&  parameter ($"$/kpc) &[\msun\ yr$^{-1}$]&$N2$,$O3N2$,$up$,$low$& &mag\\
		\hline
		a & 01:52:27.7& $-$20:00:40.8 & $-78$  & $11.5"$ / 60      & $1.27\pm0.03$&8.57,8.47,8.40,8.19&$8.48\pm0.09$ & $-19.95$\\
		b & 01:52:27.3& $-$20:00:51.0 & $-590$ & $15.4"$ / 80        &   $0.24\pm0.01$&8.51,8.44,8.54,8.03&$8.50\pm0.07$ & $-18.53$\\
		c & 01:52:26.8& $-$20:00:51.2 & $-325$ & $16.5"$ / 86        &  $0.16\pm0.01$&8.48,8.39,8.16,8.44&$8.44\pm0.06$ &$-18.66$\\
		d & 01:52:27.0& $-$20:01:34.4 & $-338$ & $23.7"$ / 123      &   $0.36\pm0.02$&8.63,8.68,8.78,7.66&$8.70\pm0.08$&$-19.61$\\
		e & 01:52:28.3& $-$20:01:35.2 & $-350$ & $33.1"$ / 172     & $<0.02$& --& -- &$-20.40$\\
		f & 01:52:25.1& $-$20:01:31.4 & $-316$ &  $39.4"$ / 205    &   $0.06\pm0.01$&$<$8.52,$<$8.37,8.66,7.94&$7.94\pm0.13$ &$-17.88$\\		
		\hline
	\end{tabular}
	\label{tab_params_emission}
	\begin{flushleft}
		Note. The columns of the table are: (1) galaxy ID; (2,3) the sky position; (3) the velocity shift with respect to the LLS; (4) the impact parameter; (5) dust corrected SFR; (6) O/H estimates from different calibrations; (7) O/H obtained as described in Section \ref{sec_em_met}; (8) total absolute magnitude based on HST/WFPC2 $F702W$ filter.\\
		$^1$$\Delta v$ is calculated with respect to the bluest absorption component at $z=0.38296$ (Table \ref{tab_all_in_absorption}).\\
	\end{flushleft}
\end{table*}
\begin{figure}
	\centering	
	\vbox{
	\includegraphics[width=.5\textwidth,bb=18 180 594 612,clip=,angle=0]{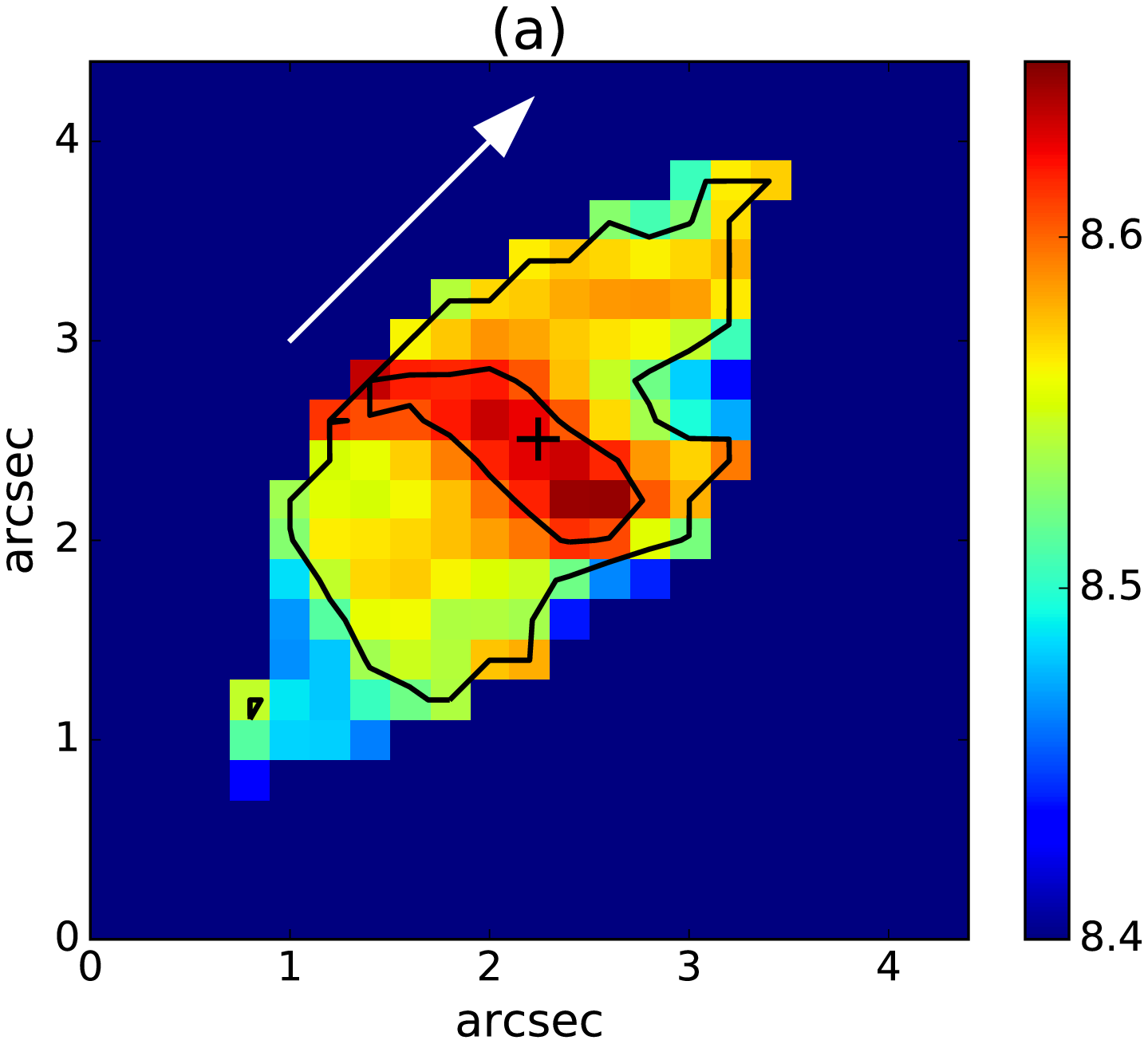}
	\includegraphics[width=.5\textwidth,bb=18 180 594 612,clip=,angle=0]{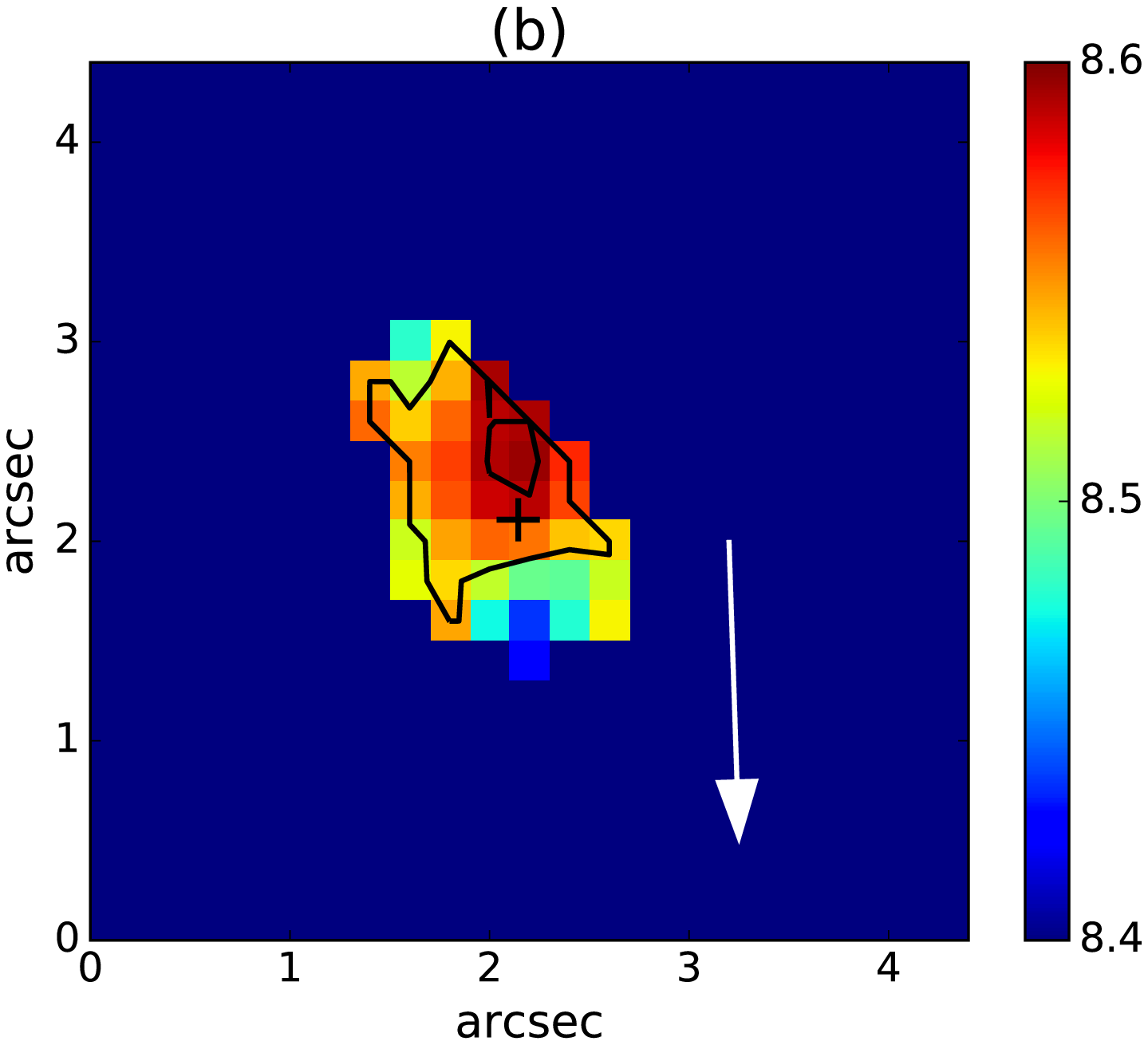}
}
	\caption{The 2D map of the emission metallicity for galaxy ``a'' and ``b'', respectively. Contours are drawn for iso-metallicity curves with 12$+{\rm [O/H]}=8.53,8.61$ for galaxy ``a'' and 12$+{\rm [O/H]}=8.53,8.59$ for galaxy ``b''. The arrow indicates the direction towards the quasar. The cross sign indicates the galactic center as determined from the stellar continuum light distribution. The mismatch between the metallicity peak and the center of galaxy ``b'' is probably due to the low detection significance of [NII]6583.}
	\label{fig_map_met}
\end{figure}
\begin{figure}
	\centering	

	\includegraphics[width=.5\textwidth,bb=18 180 594 612,clip=,angle=0]{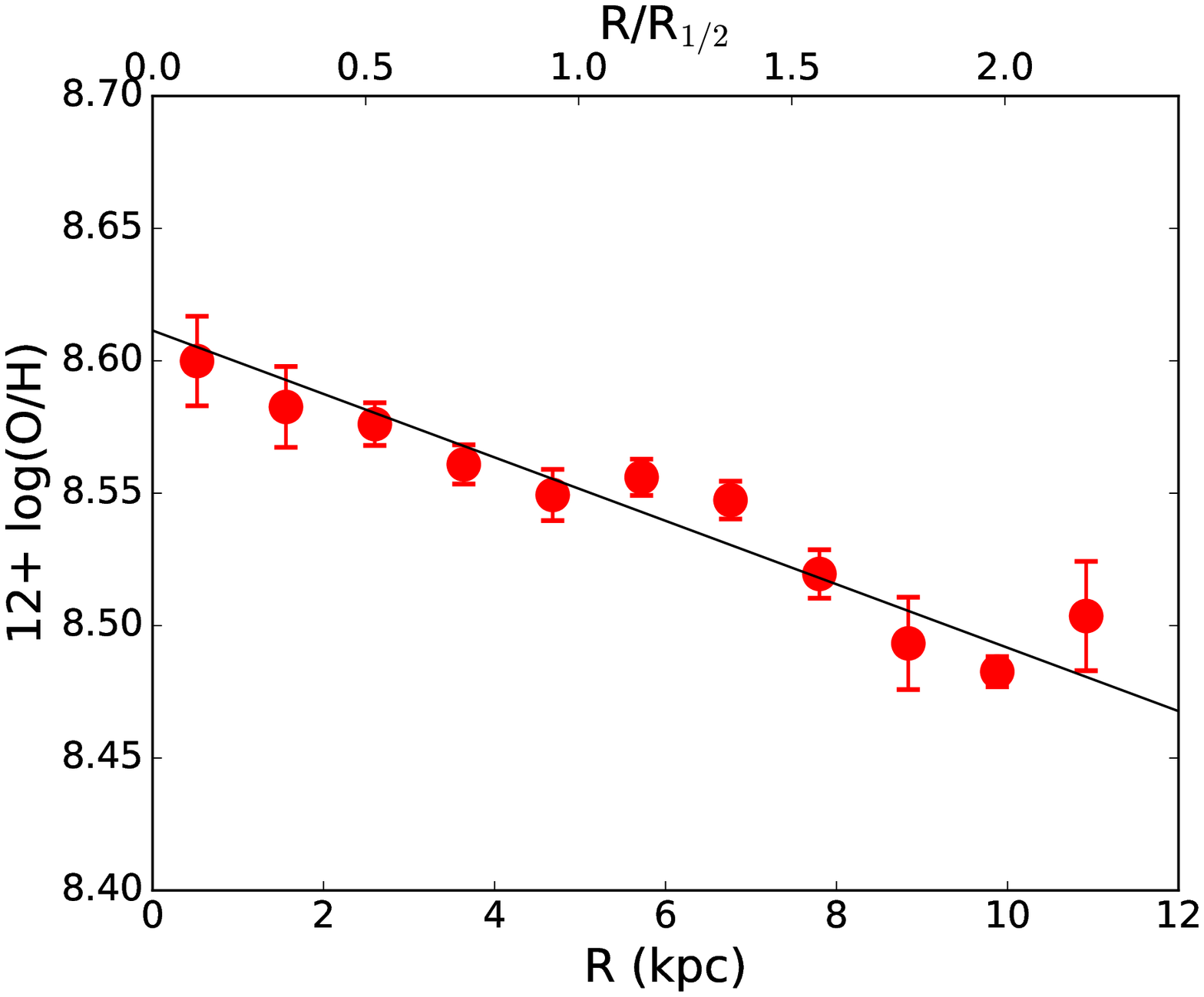}
	\caption{The 1D profile of the oxygen abundance for galaxy ``a'' where the abundances are averaged along an axis that connects the centre of galaxy ``a'' to the quasar. For the top x-axis we have assumed a half light radius of $R_{1/2}=5.4$ kpc which is based on the morphokinematic model of {$\rm H\alpha$} emission line (see Table \ref{tab_galpak}). The best linear fit, shown as a solid line, results in an O/H gradient (or equivalently a metallicity gradient) of $-0.012\pm0.003$ dex kpc$^{-1}$. The true metallicity gradient for this galaxy is a factor of two larger which is $-0.024\pm0.006$ dex kpc$^{-1}$ (see section \ref{metallicity_gradient} for more details).}
	\label{fig_grad_met}
\end{figure}
\subsection[z-em]{Metallicity of the ionized gas}\label{sec_em_met}
In this section we measure the oxygen abundance (O/H) of the \HII\ regions associated with these galaxies. We further take the oxygen abundance as an indication of the total metallicity of the ionized gas; hence we refer it to emission metallicity (Z$_{\rm em}=$(O/H)-(O/H)$_\odot$) whenever stated with respect to solar abundance \citep[12+$\log$(O/H)$_\odot$=8.69,][]{Asplund09}. 
\subsubsection{Emission metallicity measurement}
We obtain the O/H of the ionized gas using the flux ratios of different nebular emission lines based on $N2$, $O3N2$ and $R23$ indexes \citep{Pagel79,Pettini04,Kobulnicky99}. We have detected the [\NII]\,$\lambda$6583 emission line for all the emitters in this group but for galaxy ``f''. In practice, we first derive the O/H of these  galaxies using $N2$ and $O3N2$ based on a calibration given by \citet{Pettini04}. We further find the upper and lower branch values of O/H based on $R23$ as calibrated by \citet{Kobulnicky99}. Four different measurements of the O/H are summarized in the seventh column of Table \ref{tab_params_emission}. Out of the two $R23$ measures we rule out the branch that is inconsistent with those obtained from $N2$ and $O3N2$. We report the O/H and its 1$\sigma$ error as the mean of the remaining three measurements and their standard deviation, respectively. We present the final values of the O/H in the last column of Table \ref{tab_params_emission}. 

It is worth noting that our three different measurements of O/H are based on a similar set of emission line fluxes. Hence, they are not statistically independent parameters. Therefore, their measured standard deviations reflect the allowed range for O/H via systematic uncertainties introduced by different calibrations rather than being statistical errors. 

Emission metallicities of galaxies ``a'', ``b'' and ``c'' are consistent with each other within 1$\sigma$ and correspond to Z $\simeq$ 0.6\,${\rm Z}_\odot$. The largest metallicity in this group belongs to galaxy ``d'' that has a solar metallicity. Interestingly, galaxy ``d'' has also the highest amount of intrinsic dust reddening amongst the galaxies in this group. Galaxy ``f'' has a metallicity of Z $\simeq$ 0.2\,${\rm Z}_\odot$ which is the smallest measured metallicity in this group.  
\subsubsection{Metallicity gradient}\label{metallicity_gradient}
Galaxies ``a'' and ``b'' show several extended emission lines over scales of $\gtrsim1"$. This allows us to construct 2D maps for the oxygen abundance (or Z$_{\rm em}$) for these galaxies. To obtain such maps we first smooth the emission lines by a Gaussian filter of $\sigma=1$ pixel and follow the same procedure as the one we used to obtain the integrated oxygen abundance. 

Fig. \ref{fig_map_met} presents the O/H maps for galaxies ``a'' and ``b''. The contours present iso-metallicity curves for 12+[O/H]=8.53,8.61 and 8.53,8.59 for galaxy ``a'' and ``b'', respectively. The cross sign marks the center of the galaxy as determined from the stellar continuum. As can be seen from Fig. \ref{fig_map_met}, in galaxy ``a'' the maximum value of O/H is coincident with the centre of the galaxy. We also find shallow gradients along both the major and minor axes of the galaxy. However, in the case of galaxy ``b'', there isn't a symmetrical distribution of O/H and there seems to be a small offset ($\simeq 0.4" \equiv 2$\,kpc) between the centre of the stellar continuum and the location of the highest oxygen abundance. This is partly due to the weaker [\NII]\,$\lambda$6583 flux compared to the other emission lines. Such features do not exist in the O/H map obtained based on $R23$ for galaxy ``b''. Therefore, we restrict the study of the metallicity distribution to galaxy ``a''.

 We further extract the 1D profile of O/H for galaxy ``a'' along the axis that connects the centre of galaxy ``a'' to the quasar at $b=60$ kpc. To obtain this profile we average the oxygen abundances of the spaxels along the direction perpendicular to this axis. In  the case of a symmetric metallicity profile this averaging leads to twice shallower abundance profile because of dilution of the higher metallicity regions closer to the centre with the low metallicity regions at the edges. To take this effect into account we will inflate the estimated abundance gradient by the same factor. We consider the galaxy centre as the origin and average the profiles from either side of it to obtain the final profile. Fig. \ref{fig_grad_met} presents the 1D oxygen abundance profile of galaxy ``a'' as measured towards the quasar. The filled circles and bars indicate the data points and the measured standard deviation errors. Also, the best fitted line is shown using a solid line. The slope of this line translates to an O/H gradient (or equivalently a metallicity gradient) of $-0.012\pm0.003$ dex kpc$^{-1}$. Therefore, the true gradient is $-0.024\pm0.006$ dex kpc$^{-1}$.
 
 
 The metallicity gradient of $-0.024\pm0.006$ dex kpc$^{-1}$ for galaxy ``a'' is consistent with that obtained for the host galaxy of the sub-DLA studied by \citet{Peroux17}. This is also in excellent agreement with the DLA sample study of \citet{Christensen14} who found $-0.022 \pm 0.004$ dex kpc$^{-1}$ in a study of 12 confirmed DLA galaxies. Furthermore, this slope is consistent with the average metallicity difference in emission and absorption phases reported by \citet{Rahmani16} for DLA-galaxies. This suggests that neutral gas abundances are good tracers for the metallicity of galaxies \citep[see also][]{Peroux14,Christensen14}. Furthermore, the value we deduce here is also consistent with the average value of metallicity gradients for the high luminosity sub-sample of spiral galaxies in the local Universe which is $-0.026\pm0.006$ dex kpc$^{-1}$ \citep{Ho15}. 
 
 The distribution of metals has proven to be fundamental in understanding the role of interaction, mergers and feedback in galaxy formation and evolution \citep{Cescutti07,DiMatteo09,Kewley10,Rupke10,Pilkington12,Gibson13,Barrera-Ballesteros15,Leethochawalit16}. 
 \citet{Sanchez14_grad} found a slope of $-0.1$ dex/$r_e$ for a subsample of galaxies in the CALIFA survey without clear evidence of interaction. 
 Assuming an effective radius of $r_e=5.4$ kpc (obtained from our morphokinematic study of galaxy ``a'' in Section \ref{Morphokinematic}) we obtain a metallicity gradient of $-0.13\pm0.03$ dex/$r_e$ which is in agreement with the subsample of the CALIFA survey without interactions. Therefore, galaxy ``a'' likely has not interacted much with other galaxies at similar redshifts.  
\subsubsection{Emission metallicity vs. absorption metallicity}\label{metallicity_emit_abs}
Having measured the metallicity gradient we are able to predict the emission metallicity at the sky position of the quasar and compare it with the absorption metallicity. However, before performing such a comparison, it is worth noting that the two estimates are related to two different phases of the gas in galaxies that may have gone through different enrichment histories. On the one hand, the cold neutral gas seen in absorption is the fossil record of the star formation history of a galaxy and may contain the bulk of the metals ($\sim$ 90--95\%) in star-forming galaxies. On the other hand, \HII\ regions might be self-polluted with the recently produced metals \citep{Kunth86}. \citet{Aloisi03} studied a blue compact dwarf galaxy, I Zw 18, and reported up to 1 dex lower metallicity in absorption compared to emission for the $\alpha$-elements (O, Ar and Si) and N but a consistent one for Fe. Such a difference can be attributed to the enrichment of the \HII\ regions in $\alpha$-elements by the more recent star formation. 
In a more detailed analysis, \citet{Lebouteiller06} found a consistent abundance from emission and absorption for Fe, P and Ar. \citet{Bowen05} analyzed the absorption lines associated with the DLA-galaxy towards the quasar HS 1543$+$5921 and also the emission lines from the \HII\ regions of the host DLA-galaxy SBS 1543+593. In this case a consistent metallicity is obtained over a separation of $\sim3$ kpc \citep[see also][]{Schulte-Ladbeck05}. Therefore, the metallicity estimate based on \HII\ regions can be a good estimate for the ISM metallicity. 

By extrapolating the metallicity gradient of $-0.024\pm0.006$ dex kpc$^{-1}$ to the sky position of the quasar at $b=60$ kpc we find a metallicity of Z$_{\rm em} (b=60\, {\rm kpc})=-1.5\pm0.4$ which is in agreement with the limit we obtained on the absorption metallicity Z$_{\rm abs} \gtrsim -1.2$. It has been demonstrated that the metallicity distribution of galaxies in the local Universe exhibits shallower slope at radii larger that $R_{1/2}$ \citep[e.g.][]{Sanchez14_grad}. If that also holds for galaxy ``a'' we would have extrapolated to a higher metallicity at the position of the quasar which is still not inconsistent with the absorption metallicity. We remind the reader that we have obtained the absorption and emission metallicities using two different elements of iron and oxygen, respectively. The ratio O/Fe can be quite different from solar in low metallicity galaxies. Such an effect may be negligible for galaxy ``a'' that has a high metallicity of Z$_{\rm em}$ = 0.6\,Z$_\odot$. 
\begin{table*}
	\centering
	\caption{Morphokinematic parameters of the galaxy ``a'' obtained from 3D analysis of the H$\alpha^{1}$ emission line using \textsc{galpak}.}
	\begin{tabular}{lcccccccccc}
		\hline
		Galaxy & $r_{1/2}$ & $\sin i$ & position angle & azimuthal angle      & $V_{\rm max}$ & $M_{\rm dyn}$ & $M_{\rm h}$ & $R_{vir}$ & $M_{\rm \star}$\\
		& [kpc]            &  & [degree]              &[degree] & [\kms] &  [$10^{10} M_\odot$] & [$10^{11} M_\odot$] & [kpc]&  [$10^{10} M_\odot$]\\
		\hline
		a   & $5.4\pm0.1$ & $0.98\pm0.01$  & $142\pm1$  & $10\pm1$&$148\pm2$ & $2.8\pm0.1$ & $8.9\pm0.1$ & 174 &$1.5\pm0.6$\\  		
		\hline
	\end{tabular}
\begin{flushleft}
	Note. The columns of the table are: (1) galaxy ID; (2) half light radius; (3) $\sin$ of inclination angle; (4) position angle; (5) azimuthal angle; (6) maximum rotation velocity; (7) dynamical mas; (8) halo mass; (9) virial radius; (10) total stellar mass. \\
	$^{1}$ We obtain a consistent set of parameters based on the [\OIII]\,$\lambda$5007 emission line in this galaxy using \textsc{galpak}.
\end{flushleft}
    \label{tab_galpak}
\end{table*}
\begin{figure*}
	\centering	
	\centerline{\includegraphics[width=1\textwidth,bb=-242 92 854 699,clip=,angle=0]{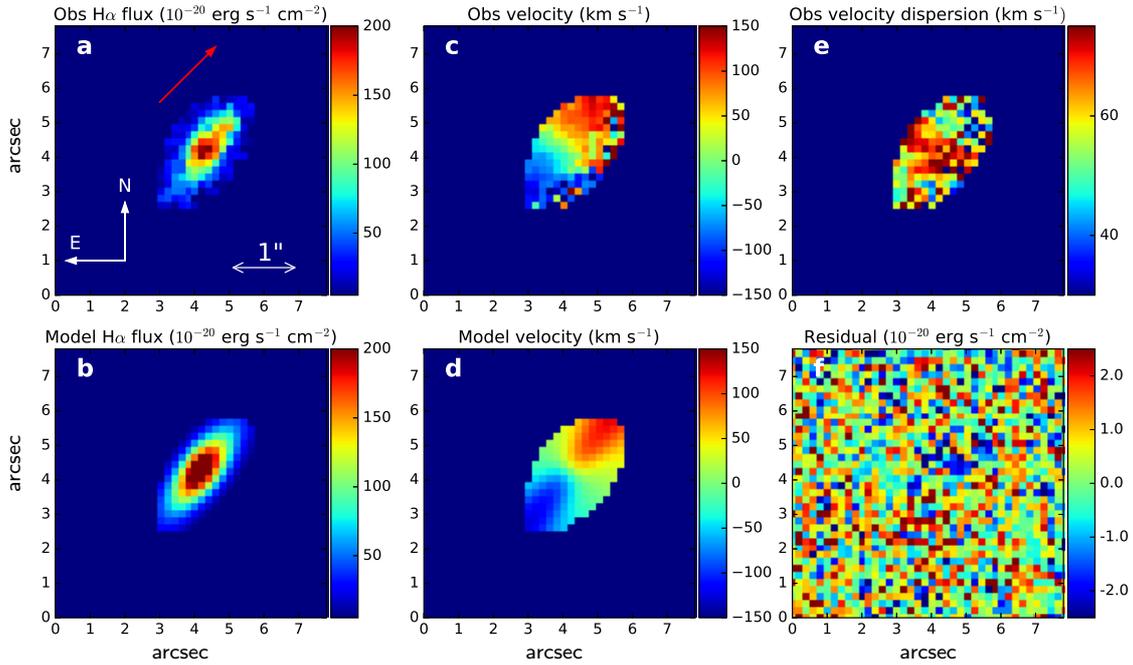}}
	\caption{Observed and modeled morphokinematic properties of galaxy ``a''. (a,b): H$\alpha$ flux map and the best obtained model based on \textsc{galpak}. The red arrow above the galaxy indicates the direction towards the quasar. The orientation of the maps is noted in the bottom left of the panel (a). Both axes are in arcsec units. The maps present a very smooth disk like distribution of H$\alpha$ emission without noticeable distortions. (c,d): Observed and modeled rotation velocity of galaxy ``a''. The modeled velocity has a maximum rotation velocity of $V_{max}=148\pm2$ \kms\ that translates to $M_{dyn}=2.8\pm0.1\times 10^{10}$ M$_\odot$ and $M_{h}=1.2\pm0.1\times 10^{12}$ M$_\odot$ (See section \ref{Morphokinematic} and eq. \ref{m_dyn} and \ref{m_halo} for the derivation of these quantities). (e): Observed velocity dispersion that peaks close to the center of the galaxy. (f): The flux residual obtained by subtraction of the best H$\alpha$ model from observed data. The parameters obtained from the model are summarized in Table \ref{tab_galpak}. }
	\label{fig_motph_g1}
	\begin{picture}(0,0)(0,0)
	\put(-155,105){ \sffamily arcsec} 
	\put(-15,105){ \sffamily arcsec}
	\put(125,105){ \sffamily arcsec}
	\put(-205,165){\rotatebox{90}{ \sffamily arcsec}}
	\put(-205,285){\rotatebox{90}{ \sffamily arcsec}}
	\end{picture}
\end{figure*}
\subsection[morphology]{Morphology of galaxies}
To study the morphology of the galaxies we carry out a similar analysis as \citet{Kacprzak07} using the HST/WFPC2 image of this field. In summary, we first fit a two component (bulge+disk) model to the light profiles of these galaxies using the \textsc{gim2d} software \citep{Simard02}. We assume a de Vaucouleurs profile and an exponential profile for the bulge and the disk of the galaxies, respectively. All the 6 members of this galaxy group are well modeled with these standard surface brightness profiles without significant residuals. Evidently, the light profiles of these galaxies do not exhibit disturbed morphologies. We further check this hypothesis by quantifying the asymmetric measure of these galaxies. We use the nonparametric $C$-$A$ method of \citet{Abraham94} which is directly calculated based on the data without modeling them. In particular we are interested in the parameter $A$ defined as:
\begin{equation}
A = \frac{\sum_{ij}(1/2)  \abs{I_{ij} - I_{ij}^{180}}}{\sum_{ij} I_{ij}} - \frac{\sum_{ij}(1/2) \abs{B_{ij} - B_{ij}^{180}}}{\sum_{ij} I_{ij}}
\end{equation}
where $I_{ij}$ and $I_{ij}^{180}$ are both counts at the pixel $ij$ of the galaxy where the former is for the original image and the latter for 180 degree rotated one. $B_{ij}$ are counts that are randomly selected from the background. The $A$ values for all these galaxies are smaller than 0.1 with the exception of galaxy ``c'' which has $A=0.14$. Such values of $A$ are consistent with the average population of normal galaxies \citep{Pawlik16} though galaxy ``c'' has a higher value consistent with the average value from host galaxies of \MgII\ absorbers \citep[e.g.][]{Kacprzak07}. In summary we do not find signatures of perturbed morphology in this galaxy group which may reflect the lack of recent strong gravitational interactions between the group members.  
\subsection[]{Morphokinematics of galaxies}\label{Morphokinematic}
To study the kinematics of the galaxies we use two different tools that are complementary to each other. At first we use the \textsc{camel} \citep{Epinat09} code to obtain the map of different emission lines. \textsc{camel} is a Python-based code that models several emission lines simultaneously in a 3D data cube. We use \textsc{camel} in a tied mode where all emission lines have the same redshift and line dispersion. Hence, velocity and dispersion maps are obtained for each galaxy. Secondly, we make use of \textsc{galpak} \citep{Bouche15} to study the morphokinematic properties of the galaxies \citep[e.g.][]{Peroux12,Bouche13,Schroetter15,Bouche16,Peroux16}. \textsc{galpak} is a Python-based tool that uses a Monte Carlo Markov Chain (MCMC) algorithm to simultaneously model the flux and velocity distribution of an emission line including 10 free parameters (sky position, flux, half-light radius, inclination, position angle, turnover radius, maximum velocity and velocity dispersion). In order to obtain a robust sample of the posterior probability we allow a 15000-long MCMC. We also check the convergence of the MCMC chain and inspect the possible correlation between each pair of parameters in their 2D distributions. Furthermore, we compare the best obtained model from \textsc{galpak} with that from \textsc{camel} to check if they are consistent. 

To have reliable parameter measurements from \textsc{galpak} two conditions need to be satisfied for a given emission line: (1) a minimum of 3$\sigma$ detection of the brightest spaxel; (2) a half light radius that satisfies $r_{1/2}/FWHM > 0.75$. The latter translates to $r_{1/2} > 0.53"$  ($\equiv2.6$ pixel) that converts to 2.8 kpc at $z=0.38$. Galaxy ``a'' is the only one satisfying these two conditions simultaneously. Therefore, we use the \textsc{galpak} results to estimate the morphokinematics of this galaxy. Table \ref{tab_galpak} summarizes the extracted morphokinematic parameters for galaxy ``a''.  

The best morphokinematic model obtained for galaxy ``a'' is presented in Fig. \ref{fig_motph_g1}. We do not find any significant feature in the residual map which means an exponential light profile with an $\arctan$ rotation curve provides a good description for the morphokinematic of this galaxy. For a consistency check we also extracted the best \textsc{galpak} morphokinematic parameters of this galaxy using the [\OIII]\,$\lambda$5007 emission line and find it to be in agreement with those obtained from H$\alpha$. 

Galaxy ``a'' is a highly inclined galaxy with $i=79\degree$ and rotates with a maximum rotation velocity of $148\pm2$ \kms. It is interesting that we find $i=76\degree$ and $r_{1/2}=7.0$ kpc for galaxy ``a'' based on GIM2D fit to the HST/WFPC2 image which is consistent with what we find from emission lines in MUSE. The velocity dispersion peaks close to the centre of the galaxy which is typical for large rotating disks. From the best obtained half light radius ($r_{1/2}=5.4\pm0.1$ kpc) and the maximum rotation speed ($V_{max}=148\pm2$ \kms) we estimate the enclosed mass within $r_{1/2}$ using the following equation:
\begin{equation}
M_{dyn}(r<r_{1/2}) =  r_{1/2}V_{max}^{2}/G
\label{m_dyn}
\end{equation}
to be $2.8\pm0.1\times10^{10}$ \msun. Assuming a spherical virialized collapse model \citep{Mo98}:
\begin{equation}
M_{h}=0.1 H(z)^{-1}G^{-1} V_{\rm max}^3
\label{m_halo}
\end{equation}
where $H(z)=H_0 \sqrt[2]{\Omega_\Lambda + \Omega_m(1+z)^3}$, we also find the halo mass of this galaxy to be $M_h=8.9\pm0.1\times10^{11}$ \msun. Such a halo mass translates to a virial radius of 174\,kpc. 

We measure the total stellar mass of this galaxy using the existence of a tight correlation between the maximum rotation velocity and the stellar mass, known as the stellar-mass Tully-Fisher relation \citep{Tully77}. Based on the relation obtained for a sample of galaxies at $<z>\sim0.6$ \citep{Puech08} and $V_{max}=148$ \kms, obtained from the morphokinematic analysis using \textsc{galpak}, we calculate the $M_{\rm \star}=1.5\pm0.6\times10^{10} M_\odot$. Therefore, the stellar-mass to dark matter fraction for this halo is $1.7\pm0.5\times10^{-2}$. This is in agreement with the stellar to dark matter fraction obtained using the technique of ``abundance matching'' which is $\sim0.02$ \citep[e.g.,][]{Behroozi10}.

We make use of \textsc{camel} to model the emission lines of the rest of the emitting galaxies in this galaxy group. In this modeling we assume all the emission lines from a pixel belonging to a galaxy to have the same redshift and line width. Such an assumption leads to a more robust estimate of the modeled parameters. Fig. \ref{fig_g2_appendix} presents the [\OII]\,$\lambda\lambda$3726,3729, [\OIII]$\lambda$5007 and H$\alpha$ emission maps of these galaxies along with their velocity maps. For galaxy ``f'' which is only partially covered by the MUSE field of view we have indicated the border of the field using a dashed line in the bottom left panel. Typical disk velocity maps can be seen for all these galaxies. Galaxies ``b'' and ``c'' have maximum rotation velocities of $\sim$ 70 \kms. This means their dynamical masses are at least $\sim$ 4 times smaller than galaxy ``a''. Galaxy ``d'' has a maximum rotation velocity of $\sim$ 240 \kms\ which is $\sim$ 1.6 times larger than that of galaxy ``a''. Hence, galaxy ``d'' has the highest dynamical and halo mass amongst the galaxies in this group. 
%
%
\section{The physical nature of the absorbing gas}
In this section we put together all the available observations of this quasar field to discover the nature of the gas we see in absorption at $z=0.38$. 
%
\subsection{Gas associated with an overdensity}
We have detected 6 galaxies in the MUSE field of view residing around the quasar sightline close to the absorption redshift. These galaxies have impact parameters from 60 kpc to 205 kpc and velocity separations from $-78$ \kms\ to $-590$ \kms\ with respect to the absorption redshift. The dark matter halos of some of these galaxies may overlap and lead them to interact strongly. Such interactions can remove a substantial amount of gas from galaxies to form the intra-group medium \citep{Thilker04,Chynoweth08,Mihos12,Hess17}. In this section we explore the hypothesis that the absorbing gas originates from the tidal debris or from the gas left in between these galaxies.
	
We searched for possible signs of interaction between these galaxies by analyzing their morphology extracted from the high spatial resolution HST/WFPC2 image of the field. As mentioned earlier, the light profile of all galaxies in the group can be well modeled using symmetric profiles leaving no significant residuals. We also inspect the distribution of the gaseous component in these galaxies from their H$\alpha$ fluxes and rotation velocity maps as presented in Fig. \ref{fig_motph_g1} and Fig. \ref{fig_g2_appendix}. The H$\alpha$ flux and the rotation velocity in the case of galaxy ``a'' are well modeled with symmetric profiles. Similarly, we do not find any particular anomalies in the H$\alpha$ flux and velocity maps of the rest of the galaxies. Therefore, we do not detect signatures of perturbed light distributions neither in the stellar components nor in the \HII\ regions of these galaxies. Altogether, this indicates these galaxies have not been strongly interacting recently. 

A 90\% percentile for the velocity distribution of these six galaxies translates to a velocity of 303 \kms. This is $\cong10$ times larger than the $\Delta v_{90}=30$ \kms\ of the absorption lines. This indicates that the gas detected in absorption is unlikely to be coupled to the gravitational potential of this overdense region. Moreover, the cold gas absorbers are redshifted with respect to all galaxies in this overdensity (e.g., see Table \ref{tab_params_emission}). The absorption lines are redshifted at $\sim300$ \kms\ with respect to the mean velocity of the six galaxies. This is at odds with the expectation that gas associated with this overdensity should appear at velocities within the range covered by the galaxies \citep[e.g.,][]{Kacprzak10}. In summary, these arguments favor a scenario in which the absorbing gas is associated with the halo of one of the galaxies rather than being part of the material between these galaxies.

Galaxy ``a'' has the smallest distance to the quasar sightline and also the smallest velocity separation with respect to the absorption redshift. Hence, the absorption lines in the $z = 0.38$ system are most likely associated with galaxy ``a''. We further confirm that the distance of this galaxy to the quasar sightline along with the equivalent width of the \MgII\,$\lambda$2796 line are consistent with the empirical correlation found between impact parameter and the equivalent width for host galaxies of \MgII\ systems \citep[e.g.,][]{Chen10,Nielsen13}. Therefore, in the remainder of this discussion we assume the absorbing system is associated with galaxy ``a''.
\begin{figure}
	\centering
	\vbox{
	\centerline{\includegraphics[width=1.2\hsize,bb=18 180 594 612,clip=,angle=0]{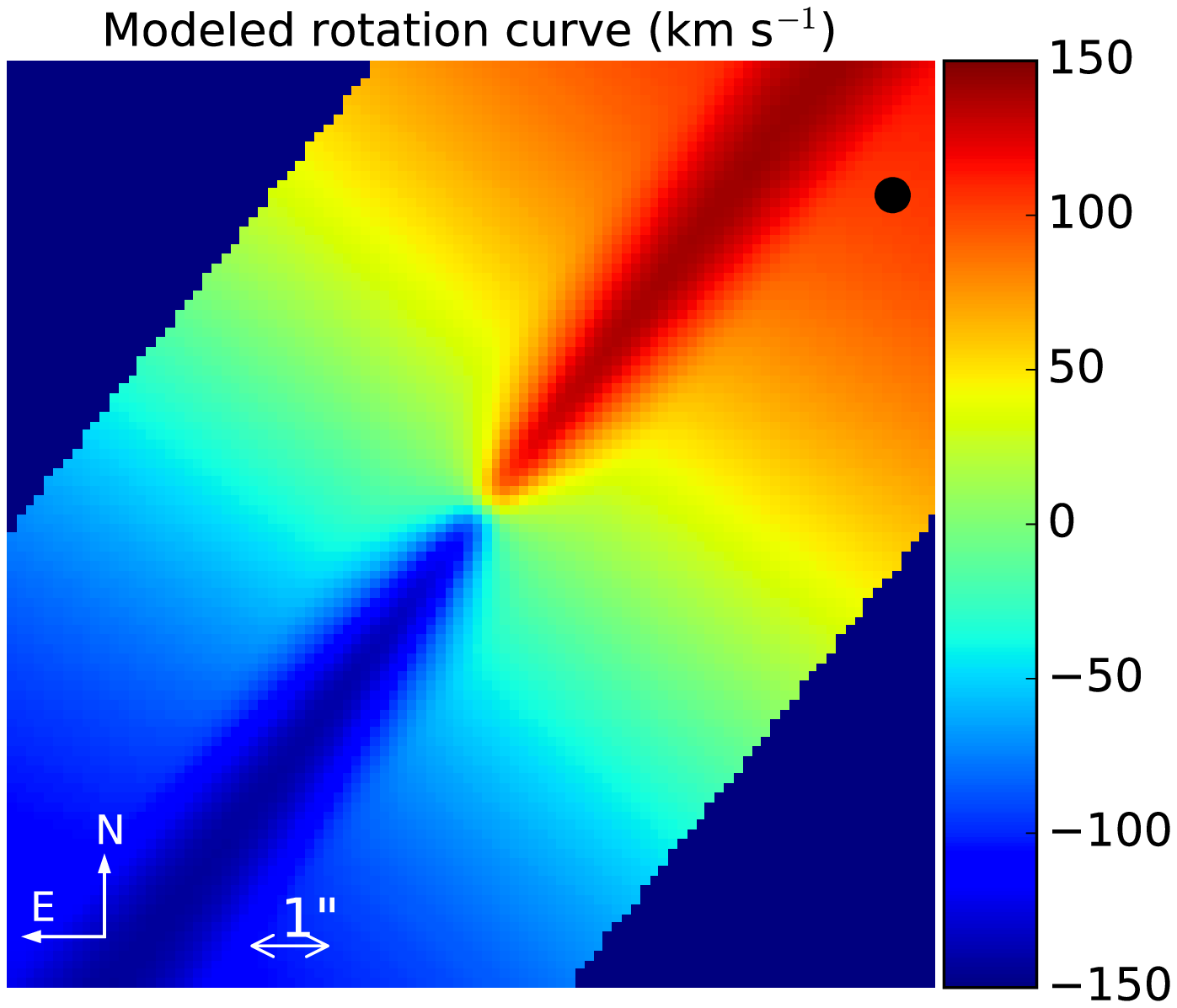}}
	\centerline{\includegraphics[width=1.\hsize,bb=93 318 518 473,clip=,angle=0]{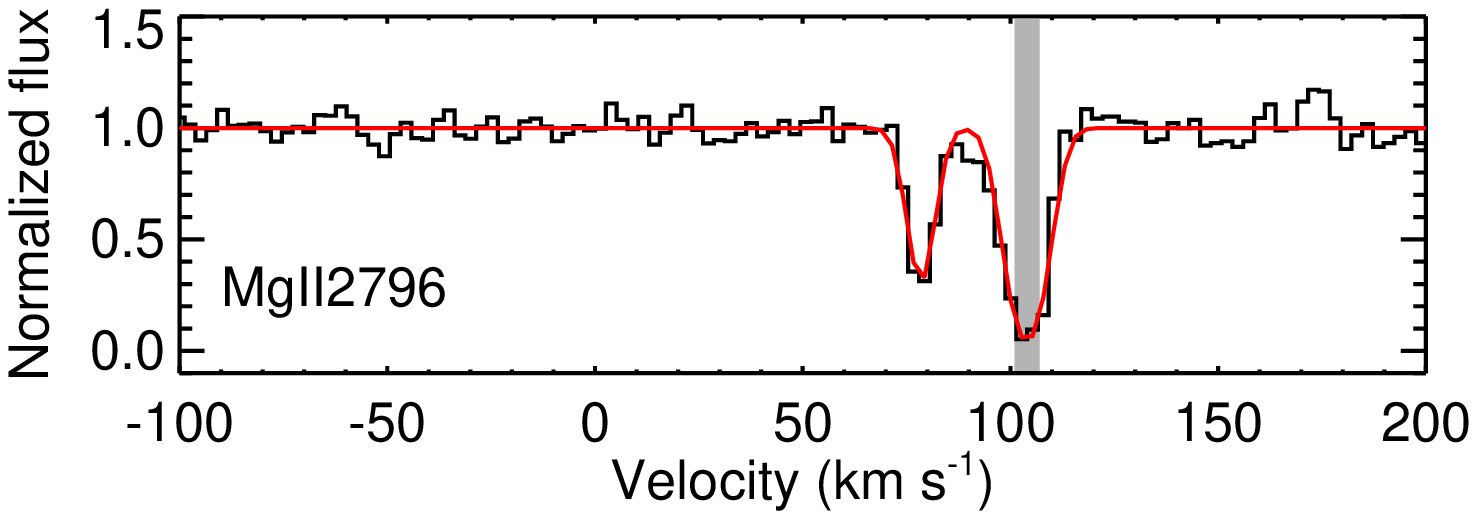}}
}
	\caption{\textit{top:} The modeled velocity of galaxy ``a'' (as discussed in Section \ref{Morphokinematic}) over an extended field including the quasar as indicated by a filled circle with a size of seeing disk, $FWHM=0.7"$ at the time of observation. The area covered by the circle spans a velocity range of 102--106 \kms. \textit{bottom:} \MgII\ absorption profile where the zero velocity is set at the systemic redshift of galaxy ``a''. The histogram and red curves are the observed and modeled profiles, respectively. The rotation velocity of galaxy ``a'' at the position of the quasar is demonstrated with the gray shaded region. Therefore, the stronger absorption component is most likely produced by the gas associated with the disk of this galaxy.}
	\label{fig_rot_gal_a}
\end{figure}
\subsection[DISK_interpret]{Galactic disk}
The gas we have detected in absorption can be kinematically associated with the disk of galaxy ``a''. This picture is motivated by the fact that the emission metallicity of galaxy ``a'', extrapolated to the position of the quasar, is consistent with the absorption metallicity (see Section \ref{metallicity_emit_abs}). However, if this picture is correct one expects the gas to corotate with the galactic disk as predicted from the rotation curve. 

To test this hypothesis we need to know the line of sight velocity of the disk at the position of the quasar, $V_{rot}^{Q}$. We extrapolate the modeled rotation curve (obtained in Section \ref{Morphokinematic}) to the position of the quasar to measure $V_{rot}^{Q}$. Fig. \ref{fig_rot_gal_a}, in the \textit{top} panel, presents the rotation curve of galaxy ``a'' over a larger field with a size of $20"\times20"$. We note that the only difference between the \textit{top} panel of Fig. \ref{fig_rot_gal_a} and panel ``d'' of Fig. \ref{fig_motph_g1} is the size of the field. The position of the quasar, as marked with a filled circle with the size of the seeing disk, $FWHM=0.7"$, corresponds to a velocity range of $V_{rot}^{Q}=$102--106 \kms. Fig. \ref{fig_rot_gal_a}, in the \textit{bottom} panel, presents the \MgII\,$\lambda2796$ absorption profile for which the zero velocity is set to the systemic redshift of galaxy ``a''. The shaded region in gray, which indicates $V_{rot}^{Q}$, matches very well with the stronger absorption component. Therefore, this absorption component most likely originates from the disk of galaxy ``a''. However, the well separated weaker absorption component at $v=78$\,\kms\ can not be explained with the rotating disk as predicted from the rotation curve.
%
%
%
\subsection{Outflow}
The ejecta from galaxies supporting outflows are usually detected to be close to the minor axis of starburst galaxies in the local Universe where the outflow penetrates the CGM perpendicular to the disk of the galaxy \citep{Heckman00,Lehnert96,Beirao15}. Hence, in quasar-galaxy pairs outflows are expected to be detected at large azimuthal angles close to 90 degrees. The azimuthal angle of the absorbing gas with respect to the major axis of the galaxy ``a'' is $10\degree\pm1\degree$. Hence, a biconical outflow from this galaxy requires a very wide cone opening angle, as large as $\sim160\degree$, to intercept the quasar sightline. Therefore, the relative geometry of this quasar-galaxy pair does not favor an outflowing scenario. Furthermore, absorption profiles produced by outflows usually have large equivalent widths \citep[e.g.][]{Tripp11Sci,Lundgren12,Muzahid15}. On the contrary, the metal absorption lines associated with this system show two components that are very narrow with a total equivalent width of $0.17$ \AA\ which is small. 

Large scale galactic winds capable of producing \MgII\ absorbers over a wide range of opening angles and impact parameters have been detected to be associated with star forming galaxies \citep{Lundgren12}. From the morphokinematic analysis using \textsc{galpak} we obtained an effective radius of $5.4\pm0.1$ kpc for galaxy ``a''. Using the integrated dust corrected SFR of $1.27\pm0.03$ \msun\ yr$^{-1}$ for this galaxy we calculate a SFR surface density of $\Sigma_{\rm SFR}=0.014\pm0.001$ \msun\ yr$^{-1}$ kpc$^{-2}$. Such a $\Sigma_{\rm SFR}$ is an order of magnitude smaller than the threshold of $\Sigma_{\rm SFR} \gtrsim0.1$ \msun\ yr$^{-1}$ kpc$^{-2}$ \citep[][]{Heckman02,Bordoloi14} required for driving the large scale galactic winds \citep[see also][]{Newman12}. In summary, the small azimuthal angle of the quasar-galaxy pair, the small equivalent width of the absorption lines and the very low $\Sigma_{\rm SFR}$ of galaxy ``a'' do not support the picture in which the absorbing system is associated with an outflow.
\subsection[fountain]{Recycled gas via galactic fountain}
Simulations suggest that the contribution of wind recycling to gas accretion onto galaxies increases with decreasing redshift \citep{van-de-Voort11b}. \citet{Christensen16} found that about half of the outflow mass across all galaxy masses is later recycled back to the host galaxies over a typical timescale of $\sim1$ Gyr. Here we explore the possibility that the absorbers may be associated with such recycled gas. 

In the most straightforward fountain model the supernova-heated ejecta travel into the halo in a radially outward trajectory up to a disk scale-height until they become thermally unstable and cool \citep{Bregman80,Houck90,Collins02,Spitoni08}. In the absence of any external torque, such a gas cloud will then fall back ballistically in a radially inward motion to reach the disk somewhere close to its point of origin. Although such a picture ignores the hydrodynamical nature of the gas, it has been able to successfully reproduce the basic features of galactic fountains \citep{Fraternali06}. In order to have the gas bubble at a large impact parameter ($b=60$ kpc) these models require large kick velocities of $v_{kick}\gtrsim250$ \kms.

To test if this LLS is associated with a fountain phenomenon we consider a supernova driven gas originating at $r\lessapprox5$ kpc with a radial velocity of $v_{out}$(=$V_{max}$) = 148\,\kms. 
The angular momentum conservation required for a ballistic motion implies that the rotational velocity of this gas will decrease to $\lessapprox12$ \kms\ in its journey to $b=60$ kpc. This velocity is 6 and 8 times smaller than the velocities of the two absorption line components. Therefore, the large specific angular momentum of the absorbing gas at $b=60$ kpc does not support such a ballistic fountain phenomenon. 


In more realistic galactic fountain models the gas particles lose their angular momentum while interacting with the hot gas in the galactic halo \citep{Fraternali08,Marinacci10}. This further enhances the discrepancy between the model predictions and the observed high specific angular momentum of the absorbing gas. Conversely, some simulations predict that the recycled materials can be accreted further out in the disk with 2--3 times more angular momentum \citep{Christensen16}. However, this is still not enough to explain a factor of 6--8 discrepancy between the observed specific angular momentum and that expected from supernova driven ejecta.
\subsection{Extended warped disk}
\HI\ 21-cm observations of galaxies in the local Universe have shown that extended warped disks are commonly associated with spiral galaxies \citep[e.g.,][]{Briggs90,Garcia-Ruiz02,Heald11}. IGM cold-mode gas accretion and interactions with nearby galaxies are considered as two possible origins of the warps \citep{Sancisi08,van-der-Kruit11}. As already discussed, we do not find signatures of recent strong interactions between galaxy ``a'' and others in this group of galaxies. Therefore, we explore if this LLS is part of a warped disk originating from IGM gas accretion.
   
In the ``cold mode'' accretion scenario the inflowing gas from the IGM is expected to form an extended warped gaseous disk, sometimes referred as ``cold-flow disk'', around the central galaxy \citep{Kimm11,Pichon11,Stewart11,Shen13}. Such an extended disk is predicted to corotate with the central disk though with a velocity offset of $\sim50-100$ \kms\ with respect to the galaxy's systemic velocity \citep{Stewart11b}. A very low azimuthal angle of the absorbing gas ($\phi=10$\degree) and a velocity offset of 78 \kms, for the bluer component, favor the cold-flow disk scenario where the accreting material resides in a coplanar geometry around the central galaxy. 

Cosmological hydrodynamic simulations predict that the inflowing gas through the cold mode accretion contains 2--3 times higher specific angular momentum compared to that of the hosting dark matter halo \citep{Stewart13,Danovich15}. The specific angular momentum, $j$, of galactic halos is typically characterized by the dimensionless spin parameter as: 
\begin{equation}
\lambda\equiv\frac{j_h}{\sqrt{2}V_{vir}R_{vir}}
\label{eq_spin}
\end{equation} 
where $V_{vir}=\sqrt{GM_{vir}/R_{vir}}$ is the virial velocity which is defined as the circular velocity at the halo virial radius, $R_{vir}$. For the halo of galaxy ``a'' we have obtained $M_{vir}$ ($\equiv M_h$) = $8.9\times10^{11}$\,M$_\odot$ and $R_{vir}$ = 174\,kpc. The spin parameter of dark matter halos is well constrained in dark matter N-body simulations with a typical value of $\lambda\simeq0.04$ for a wide range of dark matter profiles and halo masses \citep{Fall80,Mo98,Bullock01,Dutton09,Ishiyama13,Zjupa17}. Using Equation \ref{eq_spin} we find the specific angular momentum of the halo to be $j_h=\sqrt{2}\lambda$\,$V_{vir}R_{vir}$ which translates to $j_h\simeq0.06$\,$V_{vir}R_{vir}=$\,1.5 Mpc\,\kms. The absorption line associated with the ``cold-flow'' disk is detected at $b=60$ kpc (or 0.34\,$R_{vir}$) with a rotation velocity of $\Delta v=78$\,\kms\ (or 0.55\,$V_{vir}$). We estimate the specific angular momentum of the ``cold-flow'' disk, that is $j_{cf}=b\times\Delta v$, to be 0.23\,$R_{vir}\times$0.55\,$V_{vir} \simeq0.18$\,$V_{vir}R_{vir}=4.7$ Mpc\,\kms. We conclude that the cold-flow disk carries $\sim3$ times larger angular momentum compared to the halo of the galaxy which is in line with predictions of cosmological hydrodynamic simulations \citep[e.g.,][]{Stewart13,Danovich15,Stewart17}. Therefore, this absorption component may originate from a ``cold-flow'' disk.
%
%
\subsection[HVC]{High velocity clouds}
High velocity clouds (HVC) in the Local Group of galaxies are detected to have similar N(\HI) and metallicities as this LLS \citep[e.g., ][]{Lehner10,Lehner12,Richter16}. They are dominantly found within 15 kpc of their host galaxies which is 4 times smaller than the impact parameter of this LLS. HVCs associated with tidal streams of LMC and also those extending from M31 to M33 can be found up to larger distances of $\lesssim50$ kpc. Given the large impact parameter of $b=60$\,kpc and also the lack of interactions between these galaxies we discard the scenario in which this LLS originates from HVCs associated with galaxy ``a''.
\section[conclusion]{Summary and conclusions}
We have studied an LLS towards Q0152$-$020 that presents two well separated narrow metal absorption lines. Based on the upper limit N(\HI)$< 10^{18.8}$\,cm$^{-2}$ and the measured column density of \FeII, we infer the metallicity of the absorbing system to be  Z $\gtrsim 0.06$\,Z$_\odot$. Using our VLT/MUSE observations, we have discovered six galaxies close to the redshift of this absorber, and residing at impact parameters in the range  $b = 60$ -- 205\,kpc. All six galaxies are blue-shifted with respect to the absorber, with relative velocities between 78 and 590\,km~s$^{-1}$. Our main findings are as follows:
\begin{itemize}
	\item The velocity dispersion of these six galaxies translates to a virial radius and mass of 477 kpc and $8\times10^{12}$ M$_\odot$, respectively. 
	\item The extinction corrected SFR of these galaxies are in the range 0.06--1.27 M$_\odot$ yr$^{-1}$. Galaxy ``e'' is only detected in the continuum stellar light; we place a stringent upper limit on its SFR of  $<0.02$ M$_\odot$ yr$^{-1}$. Using a BPT diagram we confirm the star-forming nature of the emission lines associated with the other galaxies. 
	\item We use several calibrations to estimate the oxygen abundance of the emitting galaxies to be in a range of 12+log(O/H) = 7.94--8.70. 
	\item The 2D metallicity map of galaxy ``a'' exhibits a symmetric metallicity distribution with a mild metallicity gradient of $-0.012\pm0.004$ dex kpc$^{-1}$. By extrapolating this metallicity gradient to the apparent position of the quasar we predict a metallicity of Z$_{\rm em}$(b=60\,kpc)=0.2\,Z$_\odot$ which is consistent with the absorption metallicity.
	\item By analyzing the galaxy morphologies using a high spatial resolution HST/WFPC2 image of the field we do not find asymmetric light profiles to be associated with any of these 6 galaxies.
	\item Based on a morphokinematic analysis we obtain a maximum rotation speed of $V_{max}=148$ \kms\ for galaxy ``a''. This translates to a halo mass of $1.2\times10^{12}$ \msun\ and further to a virial radius of 174 kpc. Using the Tully-Fisher relation we find the stellar mass of $M_{\rm \star}=1.5\pm0.6\times10^{10}$ \msun\ for galaxy ``a''. 
\end{itemize}

We investigated if the gas detected in absorption towards Q0152$-020$ arises from the intra-group gas. This picture is motivated by the fact that we find 6 galaxies near the absorption redshift. However, we discount this scenario due to the narrow absorption lines and the lack of strong interactions between these galaxies. We then find this absorbing system to be most likely associated with galaxy ``a''. 

We explore several possibilities for the origin of this absorption system: (i) galactic disk; (ii) outflow; (iii) recycled gas via fountain; (iv) extended warped disk. 

Based on the rotation curve of the galaxy ``a'' and the impact parameter of $b=60$ kpc if the absorbing gas is associated with the disk of galaxy ``a'' it must have a velocity offset of $\sim102$--106 \kms\ with respect to the systemic redshift of the galaxy ``a''. One of the absorption components has a velocity of $v=104$ \kms\ that matches very well the prediction from the rotation curve. Hence, this component most likely originates from the disk of galaxy ``a''. 

A very wide opening angle of $\sim160$\degree\ is required for a biconical outflow to cross this sightline as the apparent position of the quasar is almost exactly aligned with the major axis of galaxy ``a''. Furthermore, the SFR surface density of this galaxy, $\Sigma_{\rm SFR}=0.014\pm0.001$ \msun\ yr$^{-1}$, is an order of magnitude smaller than the threshold required for driving the large scale galactic winds \citep{Heckman02}. These two considerations strongly argue against the outflow scenario. We also notice that the large specific angular momentum of the gas at $b=60$ kpc contradicts the fountain scenario in which the absorbing system is associated with recycled gas originating from SN ejecta. 

The very small azimuthal angle of the apparent position of the quasar with respect to the major axis of the galaxy ``a'' and $\Delta v$ = 78\,\kms\ velocity offset with respect to the systemic redshift favor the scenario in which the weaker absorbing component is associated with a warped galactic disk. IGM cold-mode accretion is predicted to be the main origin of such disks which are also referred to as ``cold-flow'' disk. We estimated the specific angular momentum of the ``cold-flow'' disk to be $\sim3$ times that of the hosting dark matter halo. This is in line with the results of the cosmological zoom-in simulations that predict a spin parameter of ``cold-flow'' disk to be $\approx0.2$ at $\sim0.4$\,$R_{vir}$ \citep{Stewart11b,Danovich15}. There is the yet to be understood phenomenon of high velocity clouds in the Local Group that have similar N(\HI) and metallicities as this LLS \citep[e.g., ][]{Lehner10,Lehner12,Richter16}. However, we consider this interpretation less likely as HVC clouds are observed, in the extreme cases, at distances of $b\lesssim50$ kpc, in interacting systems.

As the result of this study we have uncovered the origin of an LLS in the CGM of a galaxy at $z=0.38$. This was possible thanks to the power of IFU observation of this quasar-galaxy pair. The combination of the MUSE observations, high resolution observations of the spectrum of
the QSO and the HST image of the field provide a powerful tool to aid in disentangling the nature of the gas detected in absorption. This combination should facilitate a deeper understanding of gas flows around galaxies when the geometry is favorable.
\section*{Acknowledgements}
This work has been carried out thanks to the support of the OCEVU Labex (ANR-11-LABX-0060) and the A*MIDEX project (ANR-11-IDEX-0001-02) funded by the ``Investissements d'Avenir'' French government program managed by the ANR. HR and CP thank the ESO science visitor programme for support. This research was supported by the DFG cluster of excellence `Origin and Structure of the Universe' (www.universe-cluster.de). We are grateful to Nicolas Bouch\'{e} for developing and distributing the \textsc{galpak$^{\rm 3D}$} and further helping us in utilizing this code. GGK acknowledges the support of the Australian Research Council through the award of a Future Fellowship (FT140100933). VPK acknowledges partial support from NASA grants NNX14AG74G, NNX17AJ26G, and NASA/STScI support for program GO 13801.






\def\aj{AJ}%
\def\actaa{Acta Astron.}%
\def\araa{ARA\&A}%
\def\apj{ApJ}%
\def\apjl{ApJ}%
\def\apjs{ApJS}%
\def\ao{Appl.~Opt.}%
\def\apss{Ap\&SS}%
\def\aap{A\&A}%
\def\aapr{A\&A~Rev.}%
\def\aaps{A\&AS}%
\def\azh{AZh}%
\def\baas{BAAS}%
\def\bac{Bull. astr. Inst. Czechosl.}%
\def\caa{Chinese Astron. Astrophys.}%
\def\cjaa{Chinese J. Astron. Astrophys.}%
\def\icarus{Icarus}%
\def\jcap{J. Cosmology Astropart. Phys.}%
\def\jrasc{JRASC}%
\def\mnras{MNRAS}%
\def\memras{MmRAS}%
\def\na{New A}%
\def\nar{New A Rev.}%
\def\pasa{PASA}%
\def\pra{Phys.~Rev.~A}%
\def\prb{Phys.~Rev.~B}%
\def\prc{Phys.~Rev.~C}%
\def\prd{Phys.~Rev.~D}%
\def\pre{Phys.~Rev.~E}%
\def\prl{Phys.~Rev.~Lett.}%
\def\pasp{PASP}%
\def\pasj{PASJ}%
\def\qjras{QJRAS}%
\def\rmxaa{Rev. Mexicana Astron. Astrofis.}%
\def\skytel{S\&T}%
\def\solphys{Sol.~Phys.}%
\def\sovast{Soviet~Ast.}%
\def\ssr{Space~Sci.~Rev.}%
\def\zap{ZAp}%
\def\nat{Nature}%
\def\iaucirc{IAU~Circ.}%
\def\aplett{Astrophys.~Lett.}%
\def\apspr{Astrophys.~Space~Phys.~Res.}%
\def\bain{Bull.~Astron.~Inst.~Netherlands}%
\def\fcp{Fund.~Cosmic~Phys.}%
\def\gca{Geochim.~Cosmochim.~Acta}%
\def\grl{Geophys.~Res.~Lett.}%
\def\jcp{J.~Chem.~Phys.}%
\def\jgr{J.~Geophys.~Res.}%
\def\jqsrt{J.~Quant.~Spec.~Radiat.~Transf.}%
\def\memsai{Mem.~Soc.~Astron.~Italiana}%
\def\nphysa{Nucl.~Phys.~A}%
\def\physrep{Phys.~Rep.}%
\def\physscr{Phys.~Scr}%
\def\planss{Planet.~Space~Sci.}%
\def\procspie{Proc.~SPIE}%
\let\astap=\aap
\let\apjlett=\apjl
\let\apjsupp=\apjs
\let\applopt=\ao
\bibliographystyle{mnras}
\bibliography{/Users/hrahmani/work/IGM/files-ref/bib.bib}



\appendix
\section{emission and velocity maps of galaxies at $z=0.38$}
\begin{figure*}
	\centering	
	\vbox{
	\centerline{\includegraphics[width=1.2\textwidth,bb=-162 288 774 504,clip=,angle=0]{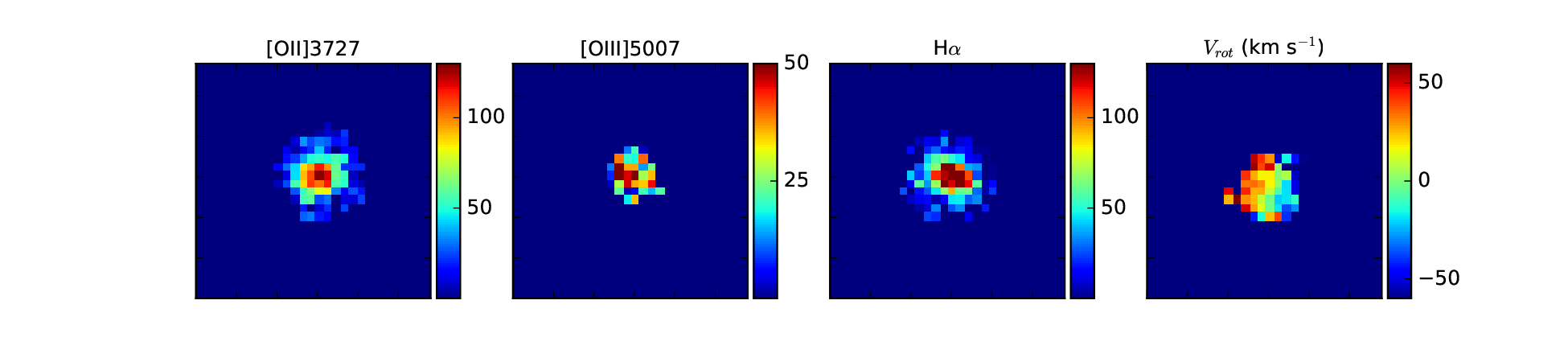}} %
	\centerline{\includegraphics[width=1.2\textwidth,bb=-162 288 774 504,clip=,angle=0]{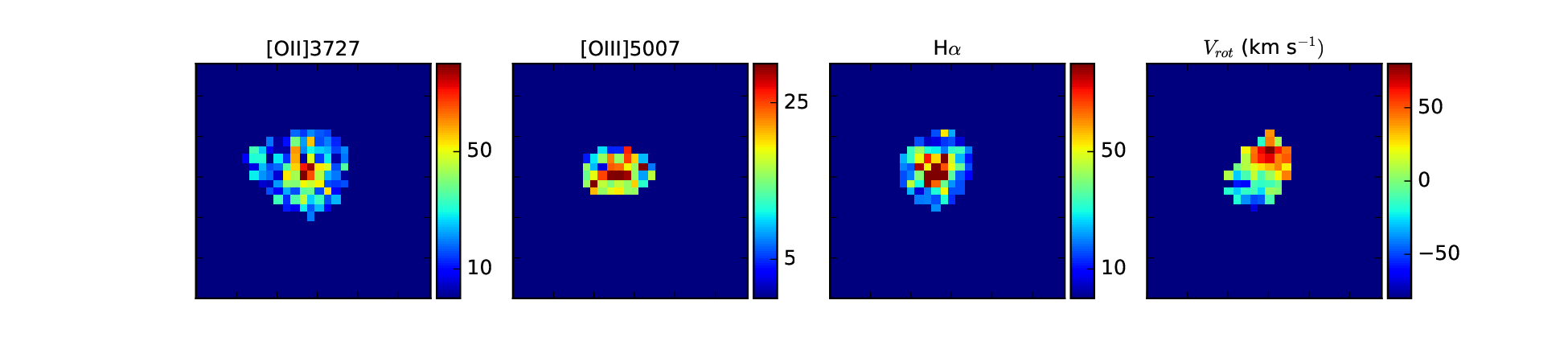}}
	\centerline{\includegraphics[width=1.2\textwidth,bb=-162 288 774 504,clip=,angle=0]{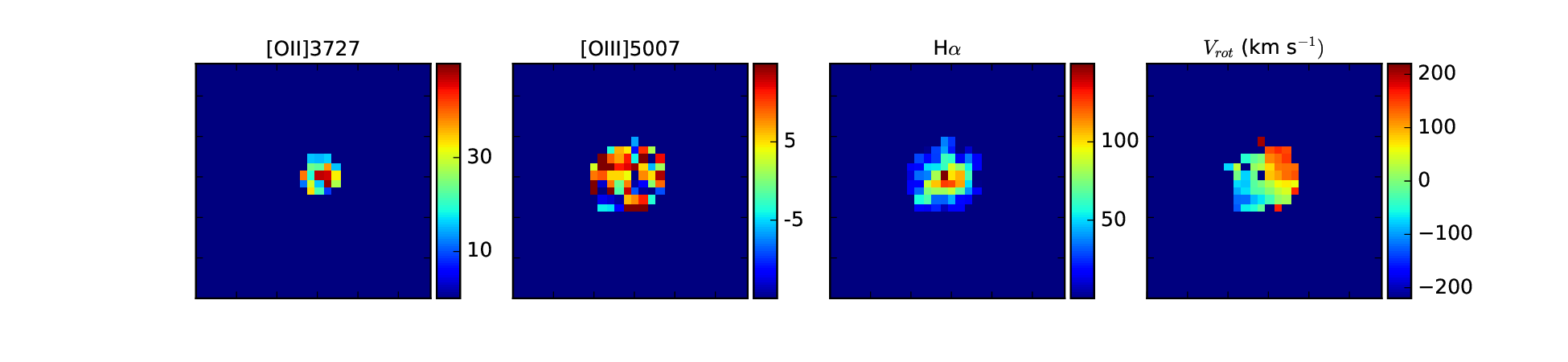}}
	\centerline{\includegraphics[width=1.2\textwidth,bb=-162 288 774 504,clip=,angle=0]{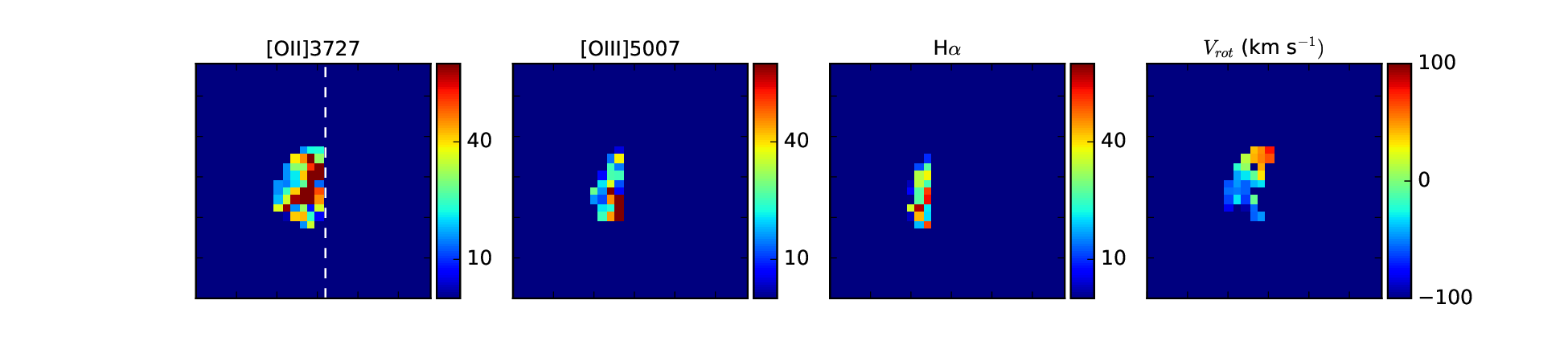}}
}
	\caption{$6"\times6"$ maps of emission lines and rotation velocity for galaxies ``b'', ``c'', ``d'' and ``f'' from top to bottom. Color-bars have units of $10^{-20}$ erg\,s$^{-1}$\,cm$^{-2}$. Arrows in the left panels show the direction towards the quasar. The dashed line in the bottom-left panel indicates the edge of the MUSE field of view.}
	\label{fig_g2_appendix}
	\begin{picture}(0,0)(0,0)
	\put(-245,530){\bf \large (b)} \put( -220,560){\textcolor{white}{\vector(0,-1){35}}}	
	\put(-245,385){\bf \large (c)} \put( -200,385){\textcolor{white}{\vector(-1,-1){25}}}	
	\put(-245,240){\bf \large (d)} \put( -220,240){\textcolor{white}{\vector(0,1){35}}}	
	\put(-245,095){\bf \large (f)}	\put( -200,110){\textcolor{white}{\vector(-1,1){25}}}		
	\end{picture}
\end{figure*}
\bsp	
\label{lastpage}
\end{document}